\documentclass{aa}
\usepackage[varg]{txfonts}
\usepackage{graphicx}
\usepackage{lscape}
\usepackage{txfonts}
\usepackage{color}
\usepackage{natbib}

\begin{document}

\newcommand\mr[1]{{{\color{mmr} #1}}}\newcommand\mrstart{\color{mmr} }\newcommand\mrstop{ \rm \normalsize \color{black} }

\definecolor{mmr}{rgb}{0.7,0.0,0.2}
\def\pre{\mr}

\newcommand\mg[1]{{{\color{mmg} #1}}}\newcommand\mgstart{\color{mmg} }\newcommand\mgstop{ \rm \normalsize \color{black} }

\definecolor{mmg}{rgb}{0.0,0.7,0.3}
\def\pre{\mg}

\newcommand\mb[1]{{{\color{mmb} #1}}}\newcommand\mbstart{\color{mmb} }\newcommand\mbstop{ \rm \normalsize \color{black} }

\definecolor{mmb}{rgb}{0.2,0.0,0.7}
\def\pre{\mb}

\title{Surface brightness-colour relations of dwarf stars from detached eclipsing binaries - I.  Calibrating sample}
\titlerunning{}
\author{D.~Graczyk\inst{1},
G.~Pietrzy\'nski\inst{2},
C.~Galan\inst{2},
J.~Southworth\inst{3},
W.~Gieren\inst{4},
M.~Ka{\l}uszy{\'n}ski\inst{2},
B.~Zgirski\inst{2},\\
A.~Gallenne\inst{4,5},
M.~G{\'o}rski\inst{2},
G.~Hajdu\inst{2},
P.~Karczmarek\inst{4},
P.~Kervella\inst{6},
P.F.L.~Maxted\inst{3},
N.~Nardetto\inst{7},\\
W.~Narloch\inst{4},
B.~Pilecki\inst{2},
W.~Pych\inst{2},
G.~Rojas Garcia\inst{2},
J.~Storm\inst{8},
K.~Suchomska\inst{2},
M.~Taormina\inst{2},
\and
P.~Wielg{\'o}rski\inst{2}
}
\authorrunning{D. Graczyk et al.}
\institute{Centrum Astronomiczne im. Miko{\l}aja Kopernika, Polish Academy of Sciences, Rabia{\'n}ska 8, 87-100, Toru{\'n}, Poland
\and Centrum Astronomiczne im. Miko{\l}aja Kopernika, Polish Academy of Sciences, Bartycka 18, 00-716, Warsaw, Poland
\and Astrophysics Group, Keele University, Staffordshire, ST5 5BG, UK
\and Departamento de Astronom{\'i}a, Universidad de Concepci{\'o}n, Casilla 160-C, Concepci{\'o}n, Chile
\and Unidad Mixta International Franco-Chilena de Astronom{\'i}a (CNRS UMI 3386), Departamento de Astronom{\'i}a, Universidad de Chile, Camino El Observatorio 1515, Las Condes, Santiago, Chile
\and LESIA, Observatoire de Paris, Universit\'e PSL, CNRS, Sorbonne Universit\'e, Universit\'e de Paris, 5 place Jules Janssen, 92195 Meudon, France
\and Universit\'e C$\hat{\rm o}$te d'Azur, Observatoire de la C$\hat{\rm o}$te d'Azur, CNRS, Laboratoire Lagrange, Nice, France
\and Leibniz-Institut f\"{u}r Astrophysik Potsdam, An der Sternwarte 16, 14482 Potsdam, Germany
}




\abstract{}
{Surface brightness -- colour relations (SBCRs) are very useful tools for predicting the angular diameters of stars. They offer the possibility to calculate very precise spectrophotometric distances by the eclipsing binary method or the Baade-Wesselink method. Double-lined Detached Eclipsing Binary stars (SB2 DEBs) with precisely known trigonometric parallaxes allow for a calibration of SBCRs with unprecedented precision. In order to improve such calibrations, it is important to enlarge the calibration sample of suitable eclipsing binaries with very precisely determined physical parameters.}
{We carefully chose a sample of ten SB2 DEBs in the solar neighbourhood which contain inactive main-sequence components. The components have spectral types from early A to early K. All systems have high-precision parallaxes from the {\it Gaia} mission. We analysed high precision ground- and space-based photometry simultaneously with the radial velocity curves derived from HARPS spectra. We used spectral disentangling to obtain the individual spectra of the components and used these to derive precise atmospheric parameters and chemical abundances. For almost all components, we derived precise surface temperatures and metallicities. }
{We derived absolute dimensions for 20 stars with an average precision of 0.2\% and 0.5\% for masses and radii, respectively. Three systems show slow apsidal motion. One system, HD\,32129, is most likely a triple system with a much fainter K6V companion. Also three systems contain metallic-line components and show strong enhancements of barium and ittrium.}
{The components of all systems compare well to the SBCR derived before from the detached eclipsing binary stars. With a possible exception of HD\,32129, they can be used to calibrate SBCRs with a precision better than 1\% with available {\it Gaia} DR3 parallaxes.}

\titlerunning{Analysis of ten detached eclipsing binary stars}
\authorrunning{Graczyk et al.}
\maketitle

\section{Introduction}
The purpose of this work is to increase the number of Double-lined Detached Eclipsing Binary stars (SB2 DEBs)\footnote{For the purposes of this paper, and at the request of an anonymous referee, we refer to eclipsing binaries for which the spectroscopic orbit of both components have been measured as SB2 DEBs even if no lines of a secondary component could be identified in spectra.} with very precise measurements of their geometrical, dynamical, and radiative properties. Gradually expanding compilations of such eclipsing binaries have been published over the last three decades \citep{and91,tor10,sou15} as they are a very useful tool in many areas of astrophysics. The well-known mass--luminosity relation for stars is calibrated with visual and eclipsing binary stars \citep[e.g.][]{mal07,eke15}. Empirical relations for the estimation of radii and masses of stars are usually derived from samples of stars based mostly on SB2 DEBs \citep[e.g.][]{tor10,eke18,moy18}. Detached eclipsing binaries provide near model-independent masses and radii of stars, and because of this they serve as prime source for calibrating and testing stellar evolutionary models. Specific subsamples of eclipsing binaries allow to test and calibrate the amount of core overshooting in intermediate-mass stars, albeit with conflicting results \citep[e.g.][]{Con18,val18,cla19,cos19}, and to predict stellar masses and ages \citep[e.g.][]{dBu18}. In some cases even a single eclipsing binary provides a stringent test of evolutionary models \citep[e.g.\ TZ For;][]{gal16,val17}.

Other applications of DEBs include the age determination of globular clusters \citep[e.g.][]{tho01,kal15} and open clusters \citep[e.g.][]{mei09,bav16}, and the determination of the helium content of a stellar cluster \citep{bro21}. They can also be used to establish bench stars with precise and accurate effective temperatures measured directly from the stars' angular dimaters and bolometric fluxes \citep{Max20,Max22}. Recently another important application was presented: calibration of the precise surface brightness -- colour relations (SBCRs) for main sequence stars based solely on DEBs \citep{gra17,gra21}.

The concept of the stellar surface brightness parameter $S\!$ is useful in astrophysics because it connects the stellar absolute magnitude with the stellar radius $R$ by a very simple relation \citep{wes69}. It is very convenient to express the $S\!$ parameter as a function of an intrinsic stellar colour -- this is a SBCR -- giving a powerful tool in predicting the angular diameters of stars \citep[e.g.][]{bar76,VBe99,ker04}. When the distance (or the trigonometric parallax) to a particular star is known the application of an SBCR immediately gives its radius \citep{lac77a}. Alternatively, when the radius of a star is known, an application of SBCR gives a robust distance \citep{lac77b}. The latter approach, in particular, has resulted in very precise distance determinations to the Magellanic Clouds \citep[e.g.][]{pie19,gra20}, setting the zero-point of the extragalactic distance ladder with a precision of $\sim$1\%.

Here we present a detailed analysis of ten new SB2 DEBs which can be used as additional calibrators of SBCRs. The sample was based on a list of eclipsing binary stars identified in data from the {\it Hipparcos} mission \citep{kru99}. This paper is one in a series of papers devoted to analysis of southern and equatorial DEBs useful in the calibration of SBCRs \citep{gra15,gra16,gra17,gra21}.

\section{Observations}\label{observ}

\subsection{Sample of stars}
Table~\ref{tab:basic} contains names and basic parameters of ten eclipsing binary stars selected for the present study. All systems are classified as double-lined spectroscopic binaries (SB2) with a possible exception of V362 Pav for which no lines of a secondary component could be directly detected and a sophisticated method was needed in order to derive its spectroscopic orbit. Because the systems are well-detached, close to the Sun and have no significant spot activity (with the exception of V963 Cen and QR Hya which both have small stellar spots), we included them in our sample. The magnitudes given are averages from catalogues listed in the SIMBAD/Vizier database,  after removing outliers and they represent out-of-eclipse brightness of the systems.

GW Eri (=HR 1300), UW LMi, QR Hya, V963 Cen, LX Mus and V362 Pav were discovered as variable stars during the \textit{Hipparcos} space mission \citep{per97}, classified as eclipsing binaries and given names in the General Catalogue of Variable Stars (GCVS) by \cite{kaz99}. HD~32129 was identified as an eclipsing binary by our team while inspecting photometry from the K2 mission campaigns \citep{how14}. V788 Cen (=HR 4624) was discovered to be an eclipsing binary by \cite{Cou71} and its name was given by \cite{kuk77}. V338 Vir was identified as an eclipsing binary by \cite{Kaz07} while CQ Ind was identified as an eclipsing binary by \cite{Ote04}; both systems were given variable star designations by \cite{Kaz08}.

Six of the objects in our sample have not previously been studied in detail, but four systems have been the subject of analysis in the past. GW Eri was reported to be a double-lined spectroscopic binary by \cite{BuM61} and a first spectroscopic orbit was given by \cite{AbL77}. The only combined analysis of spectroscopy and photometry of GW Eri before the current work was performed by \cite{Ver06}, but only an abstract has been published. A $V$-band light curve of V788 Cen was presented by \cite{Cou74}, showing two shallow and almost equal eclipses. \cite{And77} reported that this is an Am-type star and a double-lined spectroscopic binary. A preliminary analysis of V963 Cen and UW LMi based on Str{\"o}mgren $uvby$ photometry was presented by \cite{cla01}. Low quality light and radial velocity curves were used in an analysis of UW LMi \citep{mar04} as a case study of the expected performance of \textit{Gaia}. A higher-quality spectroscopic orbit based on CORAVEL spectrophotometric observations was published by \cite{Gri01}. For V963~Cen a study of its spin-axis orbital alignment and spectroscopic orbit was presented by \cite{syb18}.

\begin{table*}
\begin{center}
\caption{Basic data on the eclipsing binary stars studied in the current work.}
\label{tab:basic}
\begin{tabular}{lccccccc}
\hline \hline
ID &  R.A. (2000) & Dec (2000) & $\varpi_{Gaia/EDR3}$ & $B$ & $V$  & Orbital period & Spectral  \\
& &  & (mas) & (mag) & (mag) & (days) & type\tablefootmark{a} \\
\hline
GW Eri      &  04 11 36.20 & $-$20 21 22.2  &     $\!\!\!$11.747$\pm$0.037 & 5.977$\pm$0.017 &  5.800$\pm$0.014 & 3.659 &A1mA2-A8 \\
HD 32129   &  05 01 28.28 & +15 05 28.7   &      5.635$\pm$0.033  & 9.630$\pm$0.028   &9.093$\pm$0.025  & 16.41 &F5V               \\
UW LMi      & 10 43 30.20 & +28 41 09.1    &      9.670$\pm$0.026   &8.906$\pm$0.021  & 8.321$\pm$0.017  & 3.874&G0V            \\
QR Hya      & 10 56 31.15 & $-$34 33 50.2 &     $\!\!\!$10.672$\pm$0.024  &9.033$\pm$0.023 &  8.403$\pm$0.016 & 5.006 & G1V               \\
V788 Cen   & 12 08 53.80 & $-$44 19 33.6 &     $\!\!\!$10.908$\pm$0.045  &5.993$\pm$0.012&  5.743$\pm$0.012 & 4.966  &A2mA5-F2     \\
V338 Vir    & 13 11 17.41  & $-$11 06 21.3  &      3.905$\pm$0.020  &9.619$\pm$0.021 & 9.147$\pm$0.024 & 5.985 &F5V                \\
V963 Cen  & 13 18 44.36 & $-$58 16 01.3  &      8.725$\pm$0.018  &9.239$\pm$0.019  & 8.603$\pm$0.015  & 15.27&G2V                \\
LX Mus      & 13 40 11.53 & $-$74 04 45.0  &      6.966$\pm$0.016  &9.292$\pm$0.015  & 8.782$\pm$0.020 & 11.75&F5V               \\
V362 Pav  & 18 49 03.48 & $-$63 16 10.3   &      6.713$\pm$0.029  &7.587$\pm$0.012 &  7.403$\pm$0.014  & 2.748 &A2mA5-A9      \\
CQ Ind      &  21 31 03.29 & $-$50 50 48.9  &      9.011$\pm$0.022  &8.887$\pm$0.016  &8.360$\pm$0.016  & 8.974&F7V        \\
\hline
\end{tabular}
\tablefoot{\tablefoottext{a}{From SIMBAD database. Refined spectral types are given in Section~\ref{WD_results}}}
\end{center}
\end{table*}

\subsection{Photometry}\label{photo}

\subsubsection{Ground-based Str{\"o}mgren photometry}

We used Str{\"o}mgren $uvby$ photometry of UW~LMi and V963~Cen secured with the Str\"omgren Automated Telescope (SAT) at ESO, La Silla \citep{cla01}. The data for both stars were taken between February 1997 and March 1999. The photometry of UW~LMi comprises 734 differential magnitudes with respect to three comparison stars (HD 94218, HD 94426 and HD 91546) in each filter. The photometry of V963~Cen consists of 975 differential magnitudes in each filter with respect to HD 115031, HD 114250 and HD 117214. The photometry was detrended and normalised separately in each filter (see Table~\ref{tab:uwlmi}).

\subsubsection{Space-based photometry}
\label{sub:space}

GW~Eri was observed by the TESS space mission \citep{ric15} in short-cadence during sectors 5 and 31. For the analysis we chose the photometry from sector 31 because it has a smaller number of artifacts and outliers. The short-cadence data were downloaded, as in other cases, from the Mikulski Archive for Space Telescopes (MAST) archive\footnote{\texttt{https://mast.stsci.edu/portal/Mashup/Clients/Mast/\\Portal.html}} and contains 17\,272 photometric points. We used the Simple Aperture Photometry (SAP; \verb"SAP_FLUX"), and the data were detrended from instrumental long-term drifts using a third-order spline then normalised. We retained datapoints in eclipses and every tenth point outside eclipse, resulting in 4452 datapoints.

HD~32129 was within the field of campaign 13 of the K2 mission \citep{how14}, the extension of the {\it Kepler} space mission \citep{koch10}. The long-cadence normalised data were downloaded using the K2SFF portal on the MAST archive\footnote{\texttt{https://archive.stsci.edu/prepds/k2sff/}}. There are 3489 datapoints and for our analysis we used 321 points in and around the eclipses. HD~32129 was observed also by the TESS in sectors 5 (long-cadence), 32 and 43 (short-cadence). The short-cadence data from sector 32 cover only two secondary eclipses and we used in our analysis only data from sector 43. We used the Pre-search Data Conditioning SAP (PDCSAP; \verb"PDCSAP_FLUX") fluxes of HD~32129 containing 15\,698 photometric points. The light curve was detrended and most of the out-of-eclipse data were removed, leaving 2967 short-cadence datapoints.

\begin{figure}
\includegraphics[angle=0,scale=0.5]{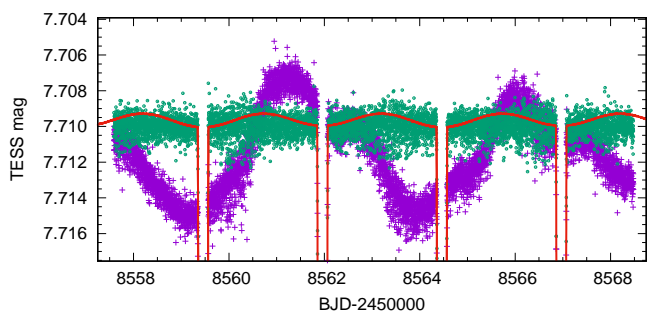}
\caption{Illustration of the detrending process for the TESS sector 9 data of QR Hya. The violet crosses show the original data, and the green circles show the detrended data. The red line is a model light curve showing the size of the out-of-eclipse proximity effects.}
\label{qrhya}
\end{figure}

QR~Hya was observed by TESS in sectors 9 and 36 in the short-cadence mode. For the analysis we used the light curve from sector 9 as it is less affected by brightness modulation due to starspots. The PDCSAP fluxes were converted into magnitudes and the light curve was detrended for the stellar activity (a modulation of $\sim$0.01\,mag) using a cubic spline -- see Fig.~\ref{qrhya}. This detrending process completely flattened the out-of-eclipse light curve, removing both the spot-modulation and the proximity effects. The latter are expected to be small ($\sim$0.001\,mag) so we decided to analyse only datapoints in the phase intervals [$-$0.05,0.05] and [0.45,0.55]. These intervals include 3137 of the original 15\,851 datapoints.

V788~Cen was observed by TESS in sectors 10 and 37 in short-cadence mode. For our analysis we used the SAP fluxes from both sectors. In the case of sector 37 we used only the second part of the light curve, as it is less affected by starspots. In our initial analysis we applied no detrending in order to retain the out-of-eclipse proximity effects. Once a satisfactory model of the system had been obtained we corrected for instrumental trends (see Section~\ref{v788cen}). We kept datapoints in eclipses and every tenth point outside of eclipses, leaving 7837 points in total.

V338~Vir was observed by the K2 in short- and long-cadence during campaign 6. However the short-cadence light curve shows a large number of instrumental drifts which proved difficult to correct. We finally used only long-cadence data, which was detrended and cleaned of outliers. In total 3309 datapoints were used in the analysis.

V963~Cen was observed by the TESS in sector 38 in short-cadence. The light curve contains 18\,495 photometric points and shows significant spot activity which affects the depth of some eclipses. After detrending and removing outliers we retained only 1010 points within and around the last two eclipses covered in sector 38.

LX~Mus was observed by TESS in two sectors 38 and 39, both in short-cadence. For our analysis we chose the PDCSAP fluxes from sector 38 because their photometric precision was higher.  The full light curve contains 17\,549 points from which we removed most of the out-of-eclipse points and ended up with 1957 datapoints.

V362~Pav was observed by TESS in sector 13 in short-cadence, giving 19\,747 datapoints. We used the SAP fluxes from the second part of the sector in our analysis. We kept all points within eclipse and every 15th point outside eclipse, resulting in a total of 2939 datapoints.

CQ~Ind was observed by TESS in short-cadence in three sectors: 1, 27 and 28. For our analysis we used the PDCSAP fluxes from sector 27. In order to follow the apsidal motion in the system we included also SAP fluxes from sector 1 covering first two eclipses. The data were detrended and normalised. We kept 3108 points within and around eclipses from sector 27, and 756 points from sector 1.

\begin{table*}
\centering
\caption{The $uvby$ photometry of UW LMi and V963 Cen.  The full data will be available at CDS.}
\label{tab:uwlmi}
\begin{tabular}{@{}ccccc@{}}
\hline \hline
Date & \multicolumn{4}{c}{Normalised Flux} \\
 HJD $-$ 2450000 & $u$ & $v$ & $b$ & $y$ \\
\hline
\multicolumn{5}{c}{UW LMi}\\
503.70509 &   0.98901&  0.98810&  0.99724&  0.99357 \\
503.70965  &  1.00647&  0.99816&  0.99816&  0.99724 \\
503.71338   & 0.99541&  1.00092&  1.00000&  1.00000 \\
503.74394  &  1.00369&  0.99357&  0.99541&  0.99632 \\
503.74853   & 1.00092&  0.99908&  1.00000&  1.00184 \\
\hline
\end{tabular}
\end{table*}

\subsection{Spectroscopy}

\label{harps}

We obtained spectra of the systems with the High Accuracy Radial velocity Planet Searcher \citep[HARPS;][]{may03} on the European Southern Observatory 3.6-m telescope in La Silla, Chile.

\begin{table*}
\begin{center}
\caption{Summary of spectroscopic observations on HARPS.}
\label{tab:harps}
\begin{tabular}{lcccc}
\hline \hline
ID & Number of spectra & Start & End & Mean S/N \\
\hline
GW Eri  & 17 &  2009 August 17 & 2021 October 25 & 110 \\
HD 32129 & 17 & 2017 December 10  & 2021 August 13 & 44 \\
UW LMi & 10 & 2017 May 27 & 2018 January 30 &  34 \\
QR Hya & 12 & 2009 February 25 & 2021 June 7 &  48 \\
V788 Cen  & 18  & 2009 February 25 & 2021 August 14 & 91 \\
V338 Vir & 18 & 2017 June 10 & 2021 August 14 & 32 \\
V963 Cen & 21 & 2009 August 17 & 2016 September 2 & 32 \\
LX Mus & 24 & 2009 February 25 & 2017 June 11 &  44 \\
V362 Pav & 21 & 2009 February 26 & 2016 August 17&  100 \\
CQ Ind & 11 & 2017 June 10 & 2021 October 24&  30 \\
\hline
\end{tabular}
\end{center}
\end{table*}

In total we collected 170 spectra between 2009 August 17 and 2021 October 25 (see Table~\ref{tab:harps}). The targets are bright and typical integration times were shorter than 10\,min; they were often used as back-up targets and also during bright sky conditions (e.g.\ near twilight). The spectra were reduced on-site using the HARPS Data Reduction Software (DRS).

\section{Analysis of spectra}
\label{analys}
\subsection{Radial velocities \label{rad_vel}}

\begin{table*}
\centering
\caption{RV measurements for eclipsing binary stars. The full data will be available at CDS.}
\label{tab_rv}
\begin{tabular}{@{}lcrcrc@{}}
\hline \hline
Object & BJD & $RV_1$ & $RV_1$ error & $RV_2$ & $RV_2$ error  \\
 & -2450000& (km s$^{-1}$) & (km s$^{-1}$) & (km s$^{-1}$) &(km s$^{-1}$) \\
\hline
UW LMi & 7901.45834 &   46.111   &0.092 &  $-$115.558&   0.097  \\
UW LMi & 7901.50942 &   42.756   &0.094&  $-$112.095&   0.098  \\
UW LMi & 7914.44475 & $-$111.967&   0.093 &   45.645&   0.097 \\
UW LMi & 7915.45642 &  $-$70.087&   0.092 &    3.063&   0.097 \\
UW LMi & 7916.44635 &   47.939 &  0.093&  $-$117.436&   0.098 \\
\hline
\end{tabular}
\end{table*}

\label{sec:rv}
We used the RaveSpan code \citep{pil17} to measure the radial velocities of the components in all systems via the broadening function (BF) formalism \citep{ruc92,ruc99}. We used templates from the library of synthetic LTE spectra by \citet{col05} matching the mean values of the estimated effective temperatures and surface gravities of the component stars. The abundances were assumed to be solar.

The line profiles of the components of HD 32129,  QR Hya,  V338~Vir, V963~Cen, LX~Mus and CQ~Ind are Gaussian and suggest small rotational velocities. The line profiles of UW LMi and V788~Cen are rotationally broadened with $v_1\sin{i}\approx v_2\sin{i}\approx20$ km~s$^{-1}$, while both components of GW~Eri rotate even faster with $v\sin{i}\approx 30$ km~s$^{-1}$.

The line profiles of the components of V362~Pav are also rotationally broadened with $v_1\sin{i}\approx40$ km~s$^{-1}$ and $v_2\sin{i}\approx20$ km~s$^{-1}$. The primary of V362~Pav is about 70 times brighter in the $V$-band than the secondary, which makes difficult to determine radial velocities of both components simultaneously. We stacked all the spectra of V362~Pav by applying radial velocity shifts to them to get a ``master'' spectrum of the primary. This spectrum was subtracted from all spectra, making the BF profile of the faint secondary much more clearly identifiable. The typical precision of an individual radial velocity measurement was about 110 m~s$^{-1}$ for the primary and 1.5 km~s$^{-1}$ for the secondary. The radial velocity measurements are summarised in Table~\ref{tab_rv}.

\subsection{Spectral disentangling}
\label{harps}
The radial velocities we derived in Section~\ref{sec:rv} were used to decompose the observed spectra of each system into the spectra of the individual components. For disentangling we used all HARPS spectra with the exception of a few spectra with a very low S/N or with very prominent solar features (when taken at bright evening/morning sky). We used the RaveSpan code which utilizes a method presented by \cite{gon06}. We ran two iterations choosing a median value for the normalization of the spectra. The disentangled spectra cover a spectral range from 4300~\AA\ up to 6900~\AA.

\subsection{Stellar atmospheric analysis}\label{abu}

\subsubsection{Methods}
\label{temp:met}

To derive the atmospheric parameters of the components of the binary systems we fitted the high-resolution (R$\sim$80000) HARPS disentangled spectra (Section\,\ref{harps}) with the `Grid Search in Stellar Parameters' {\sl GSSP} software package \citep{Tka2015}. The code uses the spectrum synthesis method by employing the {\sl S$_{YNTH}$V} LTE-based radiative transfer code \citep{Tsy1996}.  We used the {\sl LL$_{MODELS}$} grid of atmosphere models \citep{Shu2004} provided with the {\sl GSSP} code. Only the {\sl binary} mode of the {\sl GSSP} code was used to analyse the disentangled spectra. In that mode the spectra did not undergo flux renormalisation. The wavelength-dependent flux ratio $f_{\rm i}$ was calculated with the code utilizing the ratio of the components radii ($r_{1}/r_{2}$) obtained from light curve fit using the Wilson-Devinney code (see Table\,\ref{tab_par_orb}).  The {\sl binary} version does not enable the calculation of the macroturbulent velocity ($\zeta$) due to a strong correlation with the rotational velocity ($V_{\rm{rot}}$). Instead the value of $\zeta$ was estimated using published relations \citep{Sma2014, Gray2005} and held fixed at that value (Table\,\ref{T_atm_par}).

The free parameters were metallicity ([M/H]), effective temperature ($T_{\rm{eff}}$), and microturbulent velocity ($\xi$). In a few cases of objects containing metallic-lined components (GW~Eri, V788~Cen, and V362~Pav), which show strong lines mainly from ionized yttrium and barium, abundances were also calculated individually for $\sim$30 chemical elements. The abundance analysis of atmospheres of these stars will be published separately (Galan et al.\ 2022 -- in prep.). The {\sl GSSP binary} code calculates synthetic spectra for a grid of parameter values and provides the $\chi^2$ value for each pair with observed spectrum. This allowed us to judge the goodness of each fit and to choose the best-matching (corresponding to the minimum $\chi^2$) values within the grid of synthetic spectra.

Regions around the H$\alpha$, H$\beta$ and H$\gamma$ lines were excluded from the analysis. The part of the spectrum bluewards of H$\gamma$ was also excluded in most cases because it had a significantly lower S/N. The observed spectra contain in some regions the lines that have no counterparts in the line lists as well as there are the cases in the synthetic spectra that have bad data for atomic transitions. Individual masks were prepared for each object to exclude these lines from the analysis. Also, spectral regions containing artefacts from imperfectly removed features from water (H$_2$O: mainly $\lambda \sim$ 5880--6000\,\AA) and oxygen (O$_2$: $\lambda \sim$ 6274--6330\,\AA) molecules in Earth's atmosphere were skipped.

\begin{table*}[h!]
\centering
\caption{The best-fitting atmospheric parameters together with their 1$\sigma$ uncertainties estimated using the reduced $\chi^2$ and the 1$\sigma$ level in $\chi^2$ ($\chi^2_{1\sigma}$).}
\label{T_atm_par}
\begin{tabular}{@{}l|@{\hskip 1mm}l@{\hskip 2mm}l@{\hskip 2mm}l@{\hskip 2mm}l@{\hskip 2mm}l@{\hskip 1mm}|l@{\hskip 2mm}l@{\hskip 2mm}l@{\hskip 2mm}l@{\hskip 2mm}l@{\hskip 1mm}l@{}}
\hline \hline
	& \multicolumn{5}{c|}{Primary}                                                   &  \multicolumn{5}{c}{Secondary}                                                                \\
ID	& $[$M/H$]$ & $T_{\rm{eff}}$ & $\xi$  & $\zeta^{\star}$ & $V_{\rm{rot}} \sin{i}$ & $[$M/H$]$ & $T_{\rm{eff}}$ & $\xi$  & $\zeta$\tablefootmark{$\star$} & $V_{\rm{rot}} \sin{i}$ \\
        & [dex]     & [K]            & [km/s] & [km/s]          & [km/s]                 & [dex]     & [K]            & [km/s] & [km/s]                         & [km/s]                 \\
\hline
GW\,Eri	  & $+0.52$\,$\pm0.23$      & $8314$\,$\pm64$      & $4.05^{+0.28}_{-0.24}$ & $8.0$ & $24.9$\,$\pm0.8$    & $+0.58$\,$\pm0.19$                    & $8205$\,$\pm61$      & $4.34^{+0.28}_{-0.26}$      & $8.0$ & $24.0$\,$\pm0.7$    \\
HD\,32129 & $+0.19$\,$\pm0.06$      & $6713^{+76}_{-73}$   & $1.80^{+0.19}_{-0.17}$ & $6.5$ & $1.5$\,$\pm0.8$     & $+0.21$\,$\pm0.10$                    & $5777^{+145}_{-152}$ & $1.52^{+0.36}_{-0.58}$      & $3.7$ & $2.6^{+1.6}_{-2.0}$ \\
UW\,LMi   & $-0.12$\,$\pm0.07$      & $6048^{+116}_{-113}$ & $1.18$\,$\pm0.27$      & $4.0$ & $17.2$\,$\pm0.8$    & $-0.09$\,$\pm0.08$                    & $6027^{+125}_{-127}$ & $1.36^{+0.20}_{-0.28}$      & $4.0$ & $16.9$\,$\pm0.9$    \\
QR\,Hya   & $+0.00$\,$\pm0.06$      & $6012$\,$\pm64$      & $1.45^{+0.18}_{-0.23}$ & $4.0$ & $13.0$\,$\pm0.6$    & $-0.03$\,$\pm0.07$                    & $5903^{+90}_{-92}$   & $1.17$\,$\pm0.16$           & $3.5$ & $12.2$\,$\pm0.8$    \\
V788\,Cen & $+0.58$\,$\pm0.17$      & $7852$\,$\pm68$      & $4.30$\,$\pm0.25$      & $8.0$ & $20.5$\,$\pm0.7$    & $+0.36$\,$\pm0.30$                    & $7491$\,$\pm123$     & $3.36^{+0.41}_{-0.36}$      & $8.0$ & $17.3$\,$\pm1.2$    \\
V338\,Vir & $-0.07$\,$\pm0.08$      & $6723^{+135}_{-138}$ & $1.69^{+0.21}_{-0.25}$ & $6.0$ & $9.3$\,$\pm1.0$     & $-0.13$\,$\pm0.06$                    & $6464$\,$\pm62$      & $1.72^{+0.19}_{-0.17}$      & $5.5$ & $13.2$\,$\pm0.6$    \\
V963\,Cen & $-0.07$\,$\pm0.07$      & $5866^{+90}_{-85}$   & $1.36^{+0.19}_{-0.15}$ & $3.5$ & $8.2$\,$\pm0.4$     & $-0.06$\,$\pm0.06$                    & $5885^{+91}_{-88}$   & $1.41^{+0.14}_{-0.12}$      & $3.5$ & $7.9$\,$\pm0.4$     \\
LX\,Mus   & $+0.09$\,$\pm0.04$      & $6587$\,$\pm56$      & $1.69$\,$\pm0.12$      & $6.0$ & $4.0$\,$\pm0.6$     & $+0.09$\,$\pm0.04$                    & $6599$\,$\pm47$      & $1.53$\,$\pm0.12$           & $6.0$ & $4.9$\,$\pm0.5$     \\
V362\,Pav & $+0.02^{+0.15}_{-0.10}$ & $8205^{+71}_{-80}$   & $4.18$\,$\pm0.18$      & $8.0$ & $39.4$\,$\pm0.9$    & $+0.0$\tablefootmark{$\blacklozenge$} & $4900$               & $1.0$\tablefootmark{$\ast$} & $3.0$ & $19.5^{+10}_{-8}$   \\
CQ\,Ind   & $-0.01$\,$\pm0.09$      & $6524^{+138}_{-130}$ & $1.39^{+0.36}_{-0.30}$ & $5.5$ & $8.0$\,$\pm0.7$     & $-0.06$\,$\pm0.11$                    & $6224^{+180}_{-199}$ & $1.32^{+0.56}_{-0.62}$      & $4.5$ & $6.2$\,$\pm1.2$     \\
\hline
\end{tabular}
\tablefoot{\tablefoottext{$\star$}{adopted using published correlations between macroturbulence and $T_{\rm{eff}}$ or spectral type.}\\
\tablefoottext{$\blacklozenge$}{taken after the best matching model for the primary component.}\\
\tablefoottext{$\ast$}{based on the Gaia-ESO iDR6 calibration (see 3rd paragraph of Section\,\ref{temp:atm}).}}\\
\end{table*}

\subsubsection{Atmospheric parameters}
\label{temp:atm}

The input values for the parameters ($T_{\rm{eff}}$, $\log{g}$, $V_{\rm{rot}} \sin{i}$) were taken according to the results of modelling with the Wilson-Devinney code (see Section \ref{wd}, Tables\,\ref{tab_par_orb}\,and\,\ref{par_fi}). The surface gravities were not free parameters but were fixed to the values from the Wilson-Devinney code solution. Initial rotational velocities were set to the values corresponding to synchronous rotation, which is common in these types of binaries. We started to search around the solar value for metallicity $[$M/H$]$.  The input values for the microturbulent velocity ($\xi$) was estimated using published correlations with $\log{g}$, and spectral types or $T_{\rm{eff}}$ \citep{Gray2001, Gray2005, Sma2014, She2019}.

The free parameters were: $[$M/H$]$, $T_{\rm{eff}}$, $\xi$, and $V_{\rm{rot}} \sin{i}$.  The solution procedure was the same as we have used previously \citep{gra21} but now we applied only the {\sl binary} module to spectra of both components simultaneously.  We started with using relatively large steps in the grids of parameters to find the region close to the global minimum. Next, the parameter ranges were gradually narrowed and the sampling was made finer to find the solution corresponding to the best-matching model in several iterations. The 1$\sigma$ errors were estimated by finding the intersection of the 1$\sigma$ levels in $\chi^2$ ($\chi^2_{1\sigma}$) with the polynomial functions that have been fitted to the minimum values of reduced $\chi^2$ (the $\chi^2$ value normalised by the number of pixels in the spectrum minus the number of free parameters) as recommended by \citet{Tka2015}. The resulting final parameters are shown in Table\,\ref{T_atm_par}. As an example of the analysis two parts of the observed spectra for two systems -- GW~Eri and LX~Mus -- are compared with the best fit synthetic spectra in Figs.\ \ref{sp_GWEri} and \ref{sp_LXMus}.

V362\,Pav was particularly difficult to analyse compared to the other systems in this work, due to the faintness of the secondary component with respect to the primary star. Thus there was a need to limit drastically the number of free parameters for the secondary: the metallicity was fixed to that of the primary and the microturbulent velocity was set based on the Gaia-ESO iDR6 calibration (R.\ Smiljanic, 2021, private communication) as $\xi = 1.0$~km~s$^{-1}$. This $\xi$ is close to the value expected for $T_{\rm{eff}} = 4900$\,K and $[$M/H$] \approx +0.3$\,dex. The primary was reported to be a metallic star of spectral type A2mA5-A9 \citep{hou75}. Indeed, a few elements are strongly enhanced (Ba, Y, Zr) but most elements have a solar or sub-solar abundance which results in an average metallicity of only +0.02 dex.

The final temperatures are consistent with those derived from photometric colours, with an agreement generally better than 1$\sigma$ (compare Tables \ref{T_atm_par} and \ref{par_fi}), with the exception of V788~Cen which shows a slightly larger (positive) difference whilst maintaining the components' temperature ratio.  Our sample is dominated by metallicities that are near- or slightly sub-solar, with the exception of objects containing Am stars with metallicities of order $+0.5$\,dex (see\,Table\,\ref{T_atm_par}). 

\begin{figure}
\includegraphics[angle=0,scale=0.65]{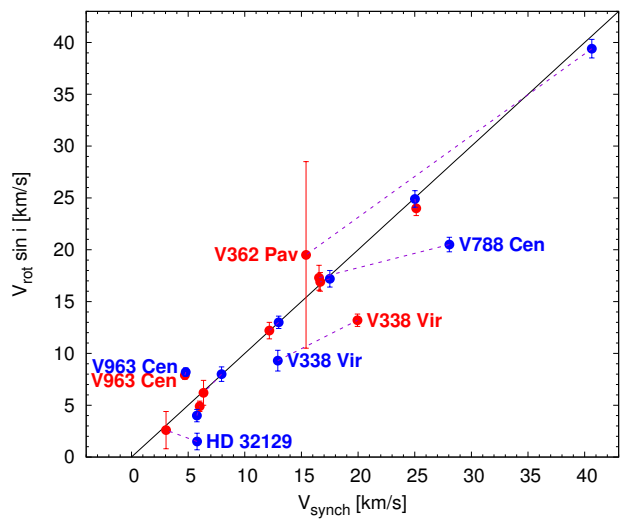}
\caption{The measured $V_{\rm{rot}} \sin{i}$ vs.\ synchronous rotation velocity for our sample.  Primary (blue) and secondary (red) components of the same system are connected by dashed lines.}
\label{f_Vrot-Vsyn}
\end{figure}

\begin{figure*}
\begin{minipage}[th]{0.73\linewidth}
\includegraphics[angle=0,scale=0.71]{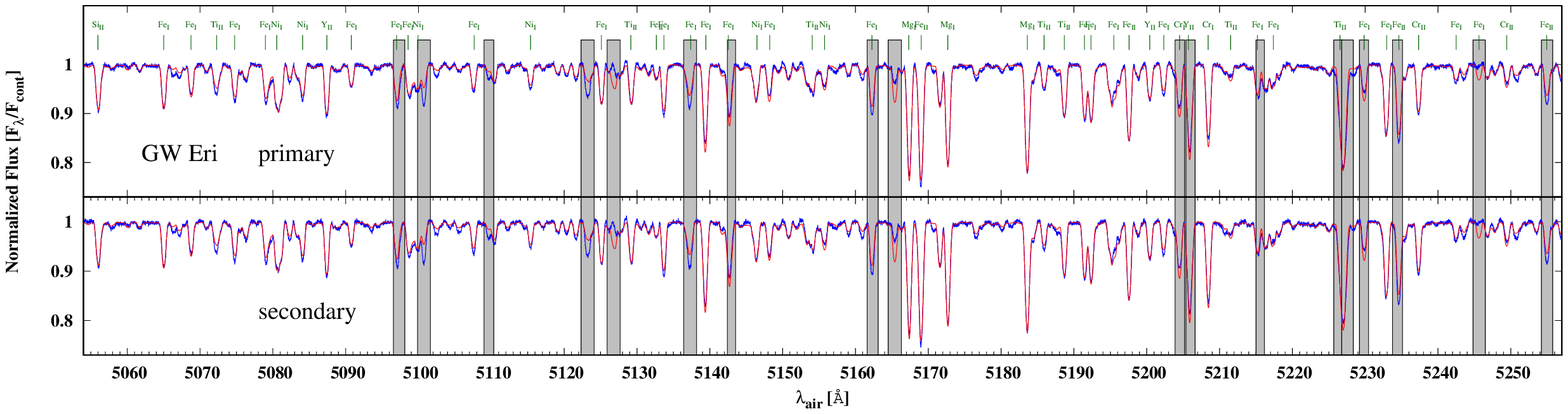}
\mbox{}
\includegraphics[angle=0,scale=0.71]{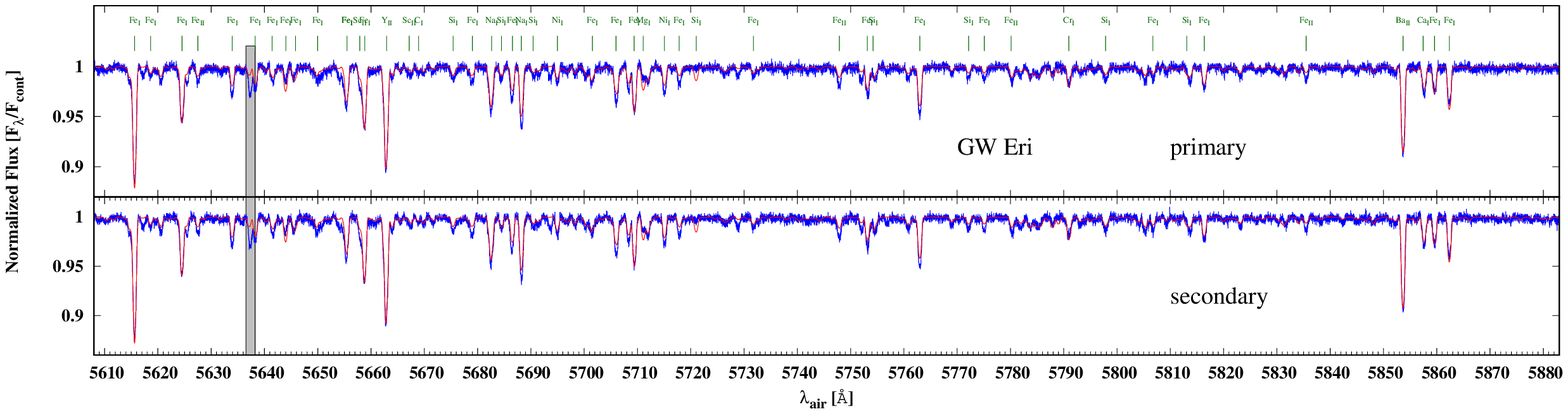}
\end{minipage}\hfill
\caption{The 5054--5257\,\AA\, (top) and 5608--5884\,\AA\, (bottom) regions of uncorrected disentangled spectra of the primary and the secondary components of GW\,Eri with selected spectral lines identified.  Strong lines from ionised yttrium (\ion{Y}{ii} $\lambda$\,5087.42, 5200.41, 5205.72, and 5662.92\,\AA), and barium (\ion{Ba}{ii} $\lambda$\,5853.67\,\AA) are visible.  The red lines denote the atmosphere model fits to the observed
spectra (blue). The gray shaded areas were excluded from calculations by the use of an individual mask.}
\label{sp_GWEri}
\end{figure*}

\begin{figure*}
\begin{minipage}[]{0.73\linewidth}
\includegraphics[angle=0,scale=0.71]{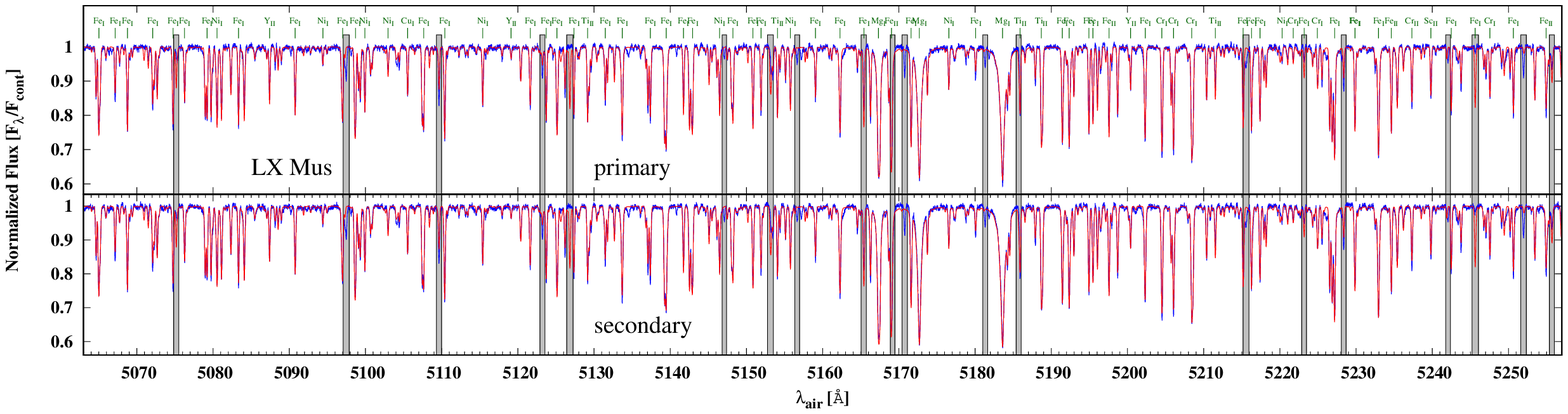}
\mbox{}
\includegraphics[angle=0,scale=0.71]{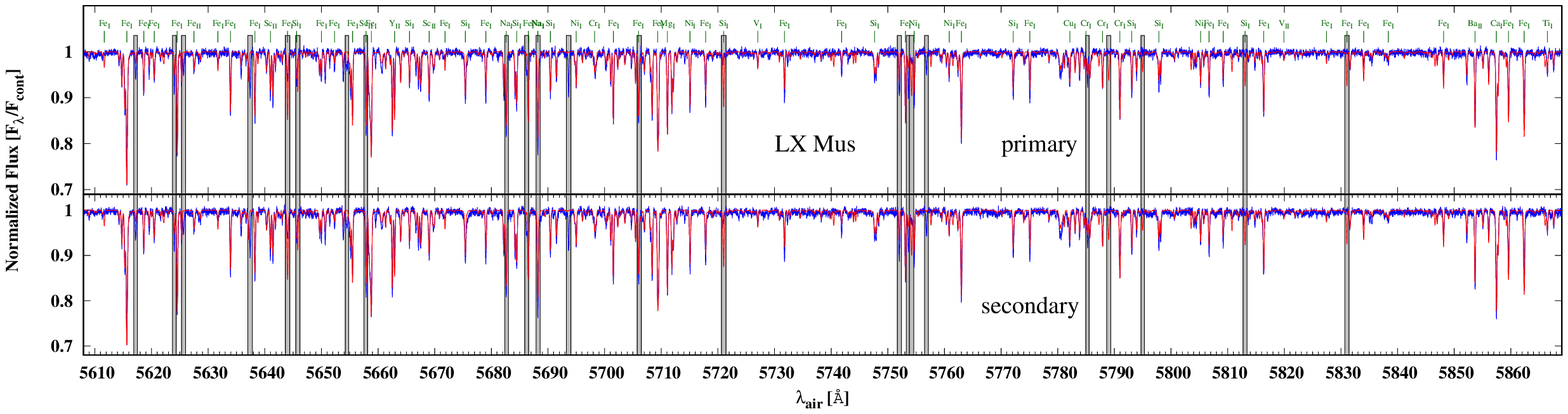}
\end{minipage}\hfill
\caption{The 5063--5257\,\AA\, (top) and 5608--5869\,\AA\, (bottom) regions of uncorrected disentangled spectra of the primary and the secondary components of LX\,Mus with selected spectral lines identified.  The red lines denote the atmosphere model fits to the observed spectra (blue). The grey-shaded areas were excluded from calculations by the use of an individual mask.}
\label{sp_LXMus}
\end{figure*}

In most cases, we found that stars rotate synchronously: their measured projected rotational velocities ($V_{\rm{rot}} \sin{i}$) are in agreement, within the 1$\sigma$ errors, with those derived from the known orbital periods and component radii (Fig.\,\ref{f_Vrot-Vsyn}). There are some oddities: the primary components of HD~32129 and V788~Cen and both components of V338~Vir rotate significantly slower and much below the synchronous velocity. A probable reason of this is that the spin and orbital axes are not aligned (small values of $\sin{i}_{\rm \,rot}$). In the case of the eccentric system V963~Cen both components rotate super-synchronously but their rotation is a factor of $\sim$2 slower than due to  synchronisation at periastron.

\section{Initial photometric analysis}

\subsection{Interstellar extinction\label{red}}

We used extinction maps \citep{sch98} with the recalibration by \cite{sch11} to determine the reddening in the direction of all ten eclipsing binaries. We followed the procedure described in detail in \cite{such15} assuming the distances from \textit{Gaia} EDR3 \citep{gaia20}. Additionally we used the three-dimensional interstellar extinction map {\sc stilism} \citep{cap17}. Finally, we adopted the average as the extinction estimate to a particular system.

\subsection{Colour -- temperature calibrations}
\label{temp:col}

To estimate the $T_{\rm eff}$ values of the eclipsing components we collected multi-band magnitudes of the systems. We use 2MASS \citep{cut03} as a good source for infrared photometry and the magnitudes were converted into appropriate photometric systems using transformation equations from \cite{bes88} and \cite{car01}. The reddening (Section~\ref{red}) and the mean Galactic interstellar extinction curve from \cite{fit07} assuming $R_V=3.1$ were combined with light ratios extrapolated from the Wilson-Devinney code (Section~\ref{wd}) in order to determine the intrinsic colours of the components. The light ratios are given in Table~\ref{tab:lratio}. We determined the $T_{\rm eff}$ values from a number of colour--temperature calibrations for a few colours: $B\!-\!V$ \citep{alo96,flo96,ram05,gon09,cas10}, $V\!-\!J$, $V\!-\!H$ \citep{ram05,gon09,cas10} and $V\!-\!K$ \citep{alo96,hou00,ram05,mas06,gon09,cas10,wor11}. For the few calibrations having metallicity terms we assumed the metallicity derived from the atmospheric analysis (see Table~\ref{T_atm_par}). The resulting temperatures were averaged for each component and are reported in Table~\ref{tab:temp}. Usually our colour temperatures are about 1$\sigma$ lower than the temperatures derived from atmospheric analysis (Section~\ref{temp:atm}). The errors reported are standard deviations of a sample of all temperatures derived for a given component. The errors include the zero-point uncertainties of the calibrations but not uncertainties introduced by disentangling of the colours.

\begin{table}
\begin{centering}
\caption{Extrapolated light ratios $l_2/l_1$ of the components. }
\label{tab:lratio}
\begin{tabular}{lccccc}
\hline \hline
 & \multicolumn{5}{c}{Photometric band} \\
System &  $B$ & $V$ & $J$ & $H$ & $K$ \\
\hline
GW~Eri & 0.9131 & 0.9281 & 0.9514 &0.9568& 0.9573 \\
HD~32129 & 0.1110 & 0.1404 & 0.2025& 0.2289 &0.2342 \\
UW~LMi & 0.8668 & 0.8751& 0.8877 & 0.8916& 0.8926 \\
QR~Hya & 0.8048 &  0.8230 & 0.8514 & 0.8603 & 0.8628 \\
V788~Cen &0.2588 & 0.2850 & 0.3321 & 0.3444 &0.3462\\
V338~Vir &2.0922 & 2.1555 & 2.2702 & 2.3036 &2.3123\\
V963~Cen &0.9753 & 0.9731 & 0.9698 & 0.9690 &0.9686\\
LX~Mus &1.1087 & 1.1005 & 1.0878 & 1.0833 &1.0832\\
V362~Pav &0.0068 & 0.0143& 0.0535 & 0.0820 & 0.0857\\
CQ~Ind &0.4714 & 0.5098 &0.5754 &0.5966& 0.6042 \\
\hline
\end{tabular}
\end{centering}
\end{table}

\begin{table}
\begin{centering}
\caption{Temperatures derived from intrinsic colours of components. }
\label{tab:temp}
\begin{tabular}{lcc}
\hline \hline
System &  \multicolumn{2}{c}{Effective temperature (K)} \\
             &  Primary   &  Secondary \\
\hline
GW~Eri & $8210\pm140$ & $8125\pm140$ \\
HD~32129 & $6705\pm59$ &$5745\pm61$  \\
UW~LMi & $6035\pm95$ & $6000\pm97$ \\
QR~Hya & $5840\pm47$ &$5760\pm50$  \\
V788~Cen & $7725\pm86$& $7225\pm67$\\
V338~Vir &$6545\pm54$&$6375\pm48$ \\
V963~Cen &$5770\pm45$&$5780\pm45$ \\
LX~Mus &$6465\pm52$& $6500\pm53$\\
V362~Pav &$8180\pm100$& $4860\pm90$\\
CQ~Ind &$6400\pm67$&$6080\pm55$ \\
\hline
\end{tabular}
\end{centering}
\end{table}

\subsubsection{Adopted values}

Precise determination of the $T_{\rm eff}$ values is very important in our approach because we did not adjust the limb darkening coefficients whilst fitting the light curves. Instead, these coefficients were automatically calculated for a given set of surface atmospheric parameters ($T_{\rm eff}$, $\log{g}$) using tables from \cite{VHa93}. Surface gravities are well determined internally within the Wilson-Devinney code, but to set the $T_{\rm eff}$ scale we needed external information. The $T_{\rm eff}$  scale was set by fixing the surface $T_{\rm eff}$ of the primary star, $T_1$, to the average of two previous $T_{\rm eff}$ determinations (Sections \ref{temp:atm} and \ref{temp:col}). The adopted $T_1$ in all cases is well within the 1$\sigma$ uncertainty of both $T_{\rm eff}$ determinations. Subsequently the $T_{\rm eff}$ of the secondary, $T_2$, was scaled according to $T_1$ during the light curve analysis with the WD code.

\section{Analysis of combined light and radial velocity curves \label{wd}}

For analysis of the eclipsing binaries we made use of the Wilson-Devinney program (WD) version 2007 \citep{wil71,wil79,wil90,van07}\footnote{\texttt{ftp://ftp.astro.ufl.edu/pub/wilson/lcdc2007/}}, equipped with a Python wrapper. When the work on the paper was well advanced we learned that a newer version of the WD code \citep[][LCDC2015, version 2019]{vHam14} included directly the TESS bandpass\footnote{\texttt{ftp://ftp.astro.ufl.edu/pub/wilson/lcdc2015/}}. We decided to use this new version which allows also for a higher grid resolution over stellar surfaces and for which a specific python GUI\footnote{\texttt{https://github.com/Varnani/pywd2015-qt5}} was written \cite[][PyWD2015]{Guz20}.

\subsection{Initial parameters}

We fixed the $T_{\rm eff}$ of the primary component during analysis to the average of the $T_{\rm eff}$ values derived from the colour--temperature calibrations and the atmospheric analysis. In all cases those two determinations are consistent to within 1$\sigma$. The standard albedo and gravity brightening coefficients for convective stellar atmospheres were chosen. The stellar atmosphere option was used (\verb"IFAT1=IFAT2=1"), radial velocity tidal corrections were automatically applied (\verb"ICOR1=ICOR2=1") and no flux-level-dependent weighting was used. We assumed synchronous rotation for both components in all systems. Both the logarithmic \citep{kli70} and square root \citep{dia92} limb-darkening laws were used, with coefficients tabulated by \cite{VHa93}.

\subsection{Fitting model parameters}
\label{sub:fitting}
With the WD binary star model we fitted simultaneously the available light curves and radial velocity curves of both components using the grid fineness parameters \verb"N1=N2"=60. In cases in which one of the stars was significantly larger than a companion it was neccesary to use a higher numerical precision and we set \verb"N"=90 for a larger star. We assumed a detached configuration in all models and a simple reflection treatment (\verb"MREF=1", \verb"NREF=1"). Each observable curve was weighted only by its {\it rms} through comparison with the calculated model curve. We adjusted the following parameters during analysis: the orbital period $P_{\rm orb}$, the epoch of the primary eclipse $T_0$ in cases of circular orbits, the phase shift $\phi$ when orbits were significantly eccentric, the semimajor axis $a$, the mass ratio $q$, both systemic radial velocities $\gamma_{1,2}$, the eccentricity $e$, the argument of periastron $\omega$, the orbital inclination $i$, the temperature of the secondary $T_2$, the modified Roche potentials $\Omega_{1,2}$ -- corresponding to the fractional radii $r_{1,2}$ -- and the luminosity parameter $L_1$. Additionally, we fitted for third light $l_3$. The best models were chosen according to their reduced $\chi^2$ and a lack of significant systematic trends in the residuals. The initial temperatures of the components were set according to their individual colours (see Section~\ref{temp:col}), then adjusted according to the results of the atmospheric analysis of disentangled HARPS spectra. Usually we took a simple mean of the colour and spectroscopic temperatures of the primary to set the temperature scale of the model. In cases when the secondary was significantly brighter then the primary we set the scale using the secondary's temperature.

\begin{figure*}
\begin{minipage}[th]{0.5\linewidth}
\vspace{-1.3cm}
\includegraphics[angle=0,scale=.46]{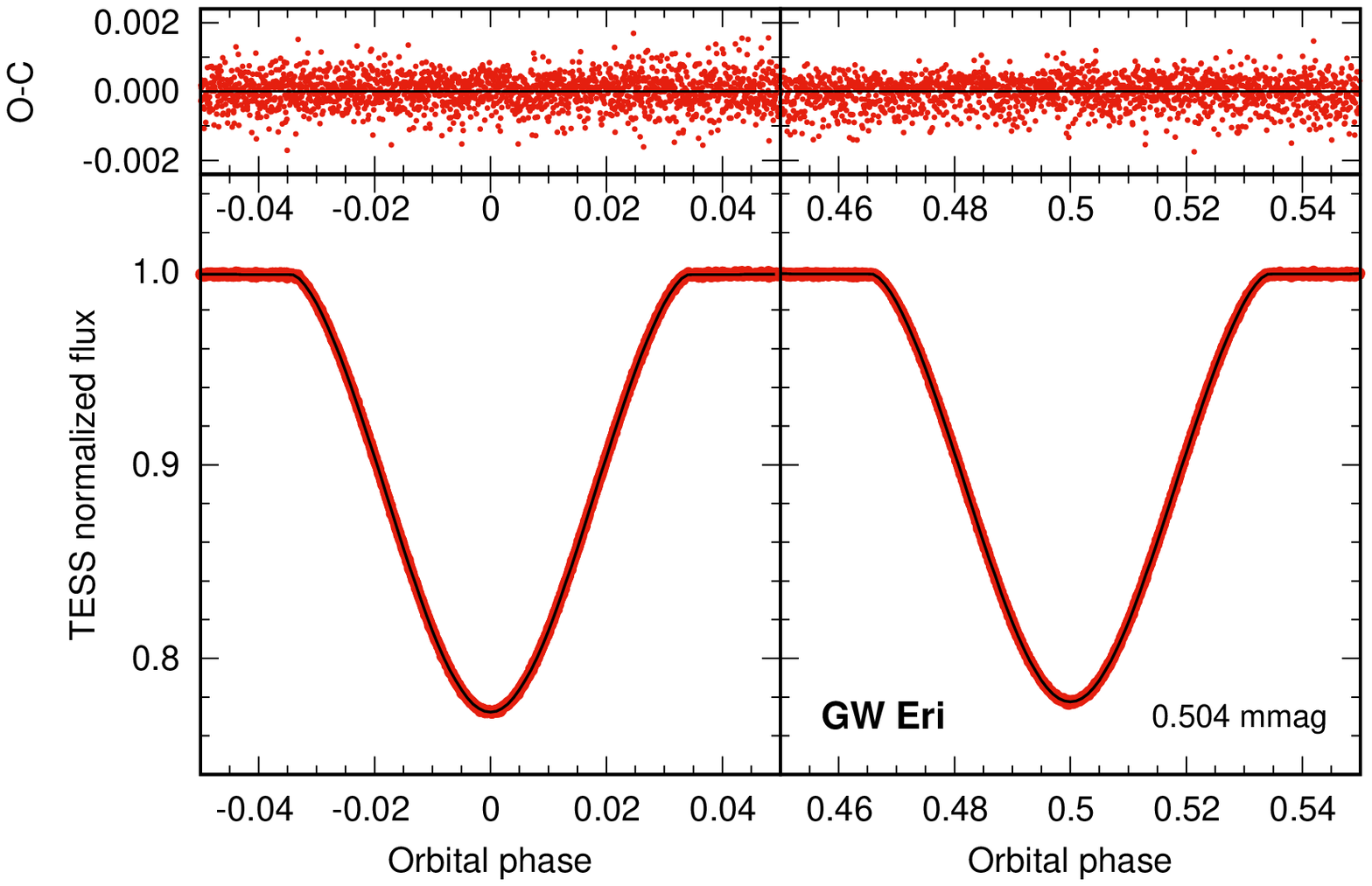} \vspace{-1.45cm}
\mbox{}
\includegraphics[angle=0,scale=.46]{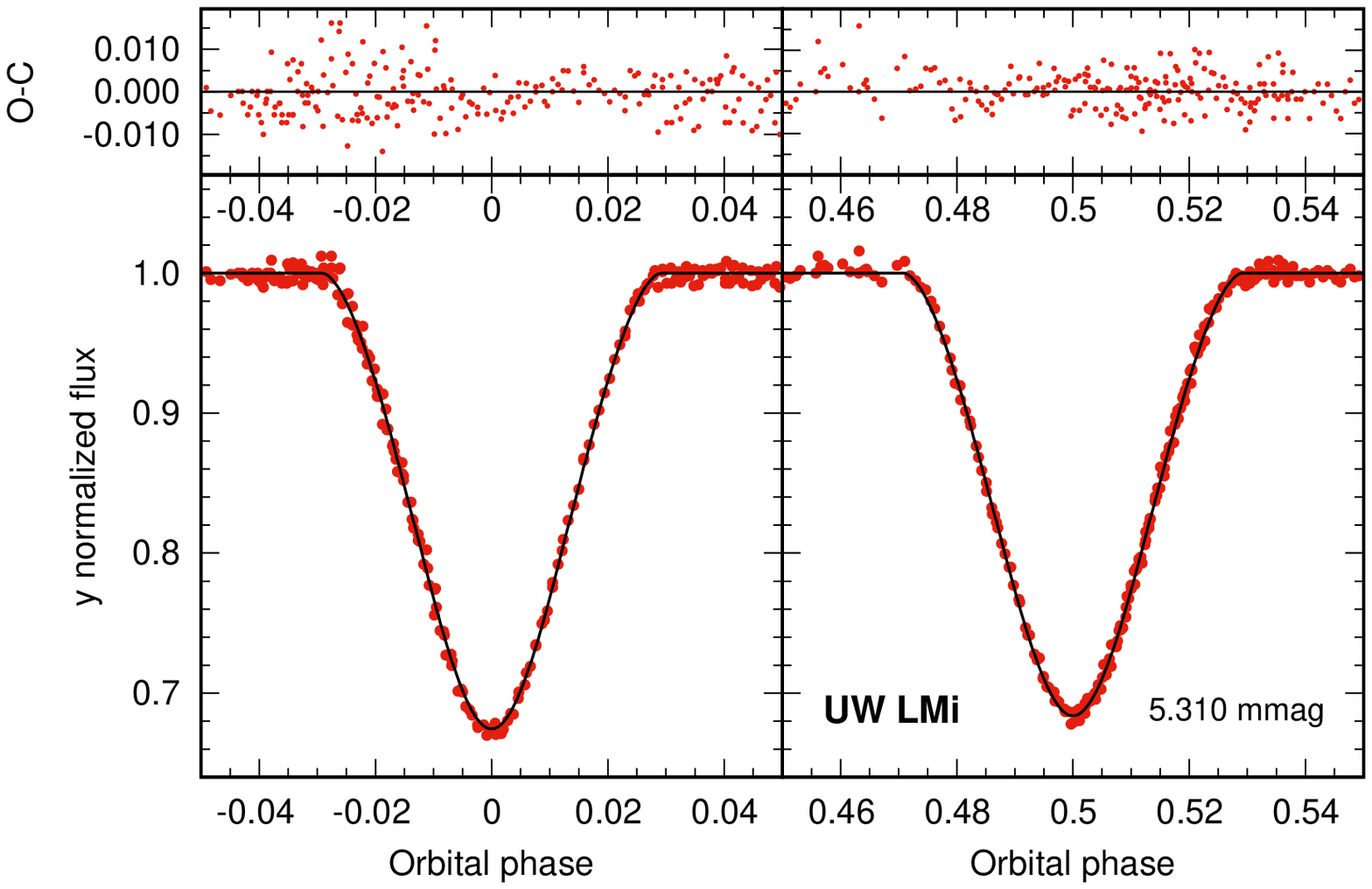} \vspace{-1.45cm}
\mbox{}
\includegraphics[angle=0,scale=.46]{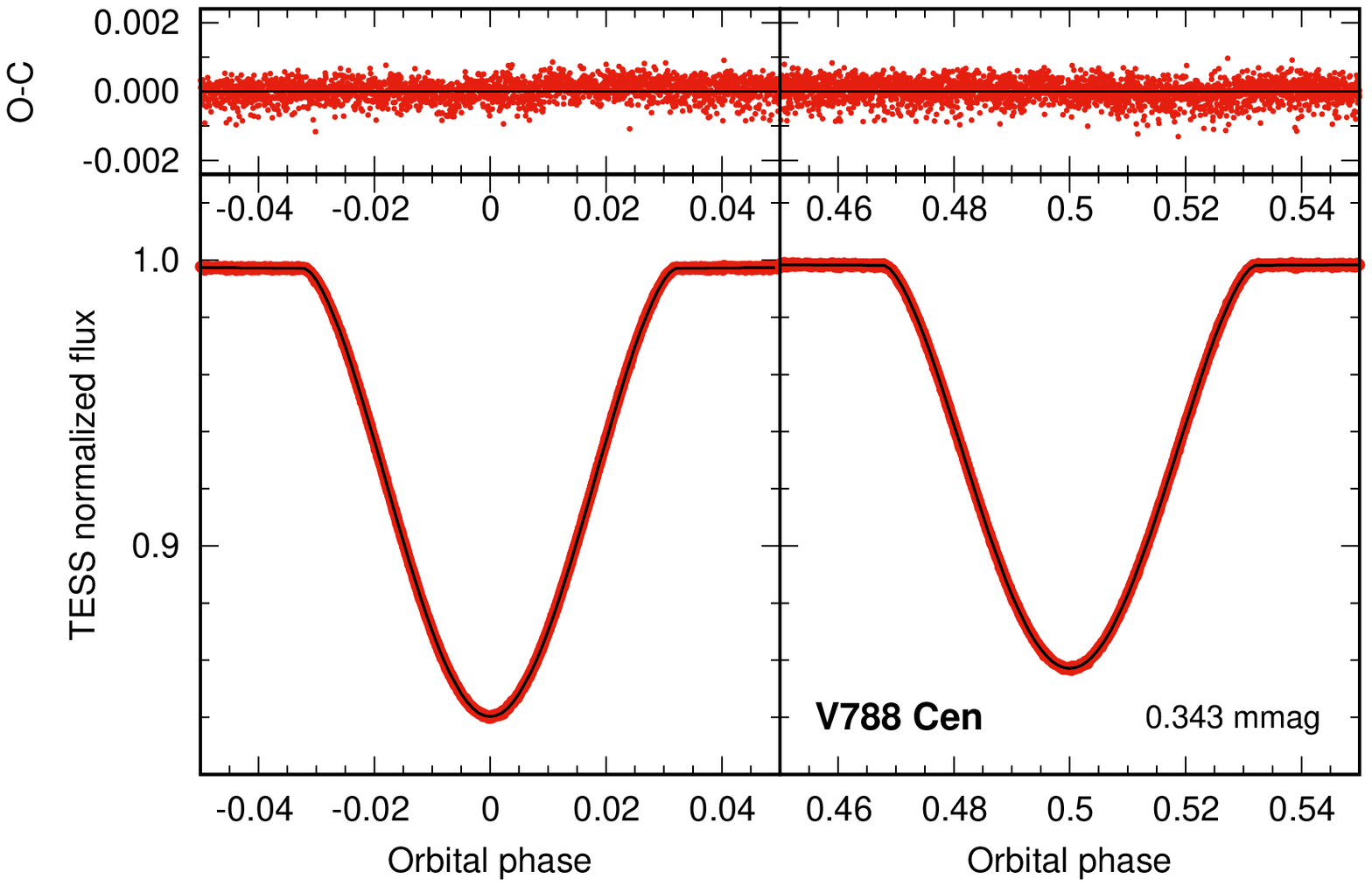}\vspace{-1.45cm}
\mbox{}
\includegraphics[angle=0,scale=.46]{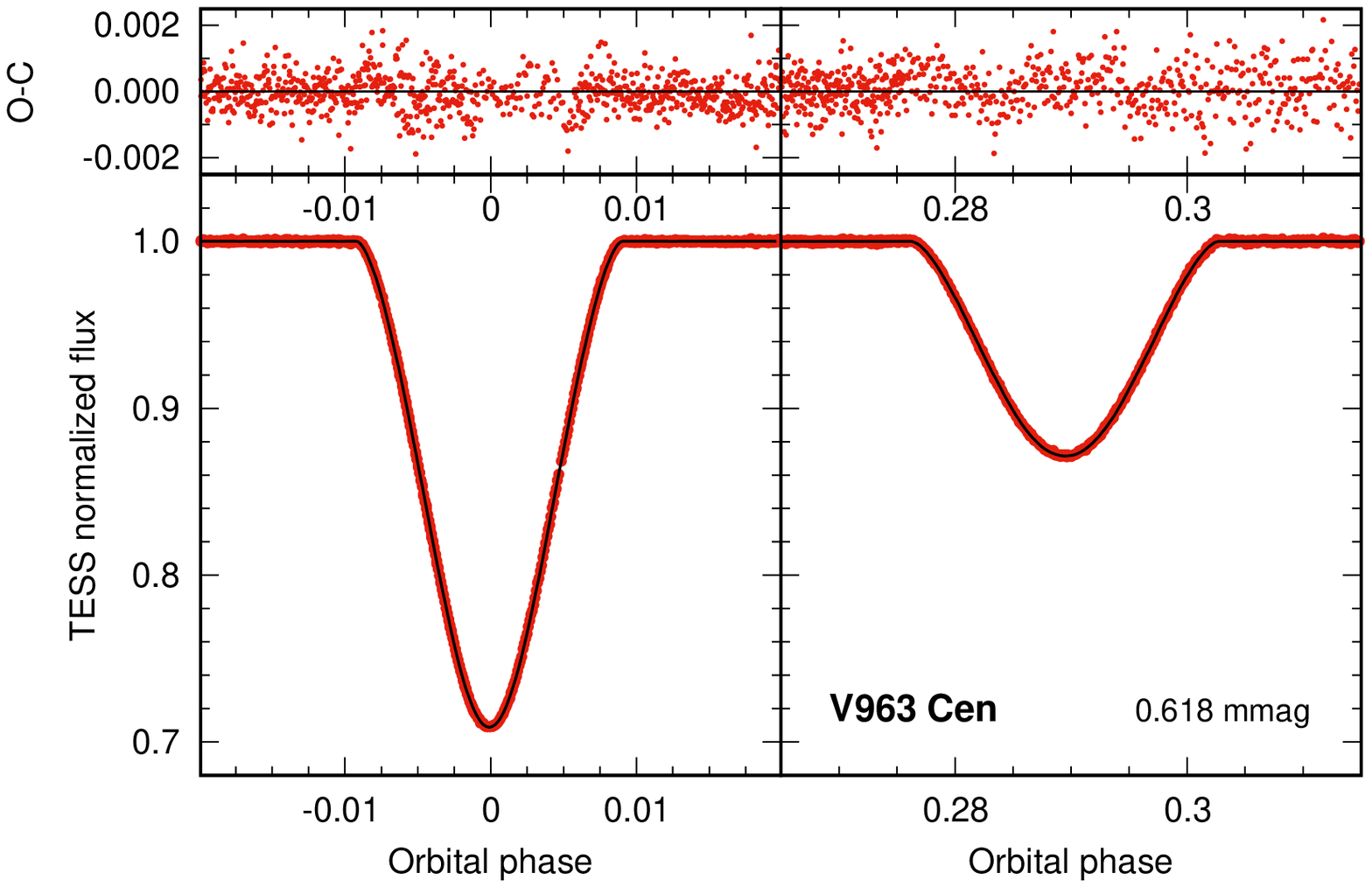}\vspace{-1.45cm}
\mbox{}
\includegraphics[angle=0,scale=.46]{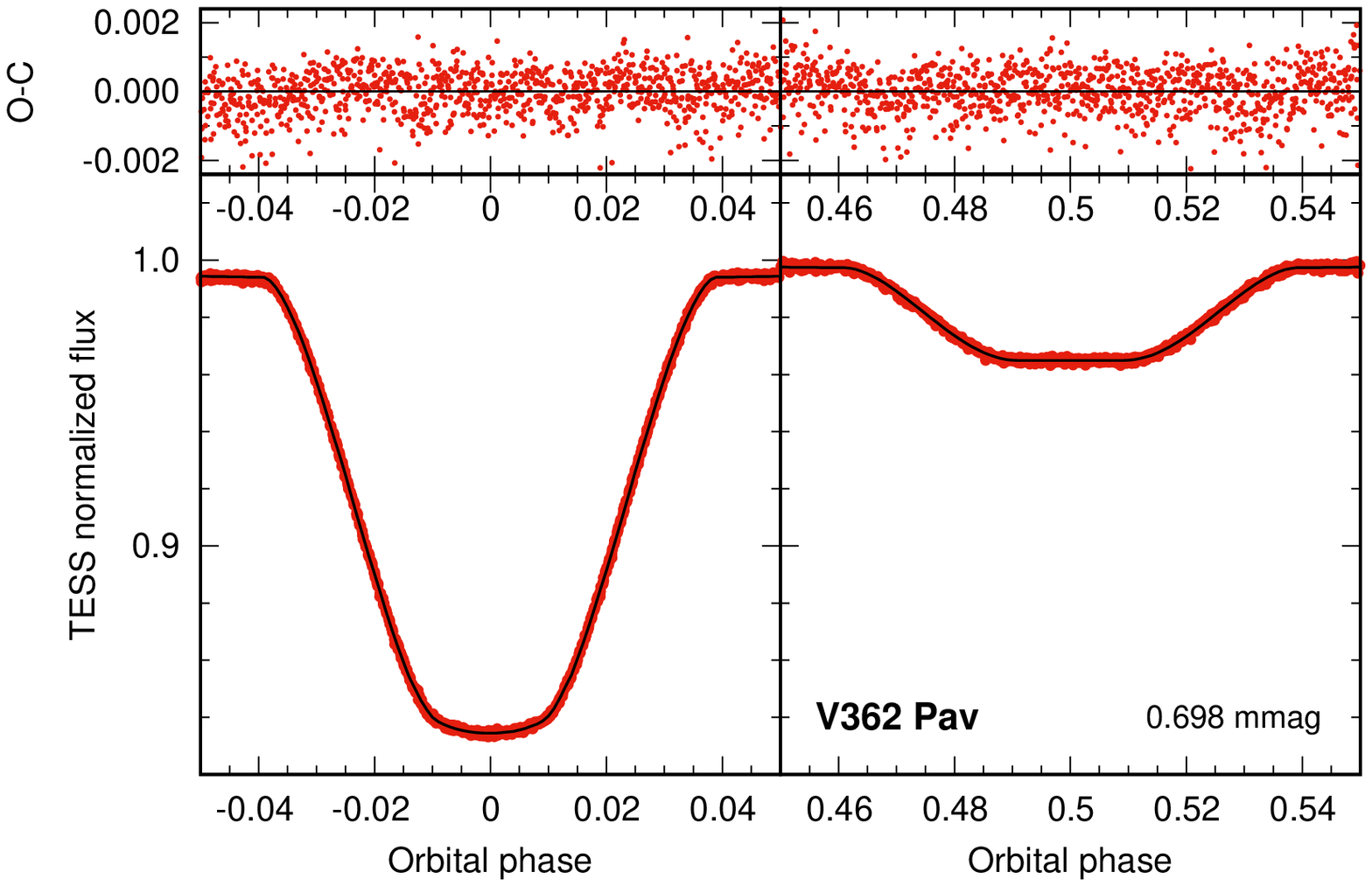}
\end{minipage}\hfill
\begin{minipage}[th]{0.5\linewidth}
\vspace{-1.3cm}
\includegraphics[angle=0,scale=.46]{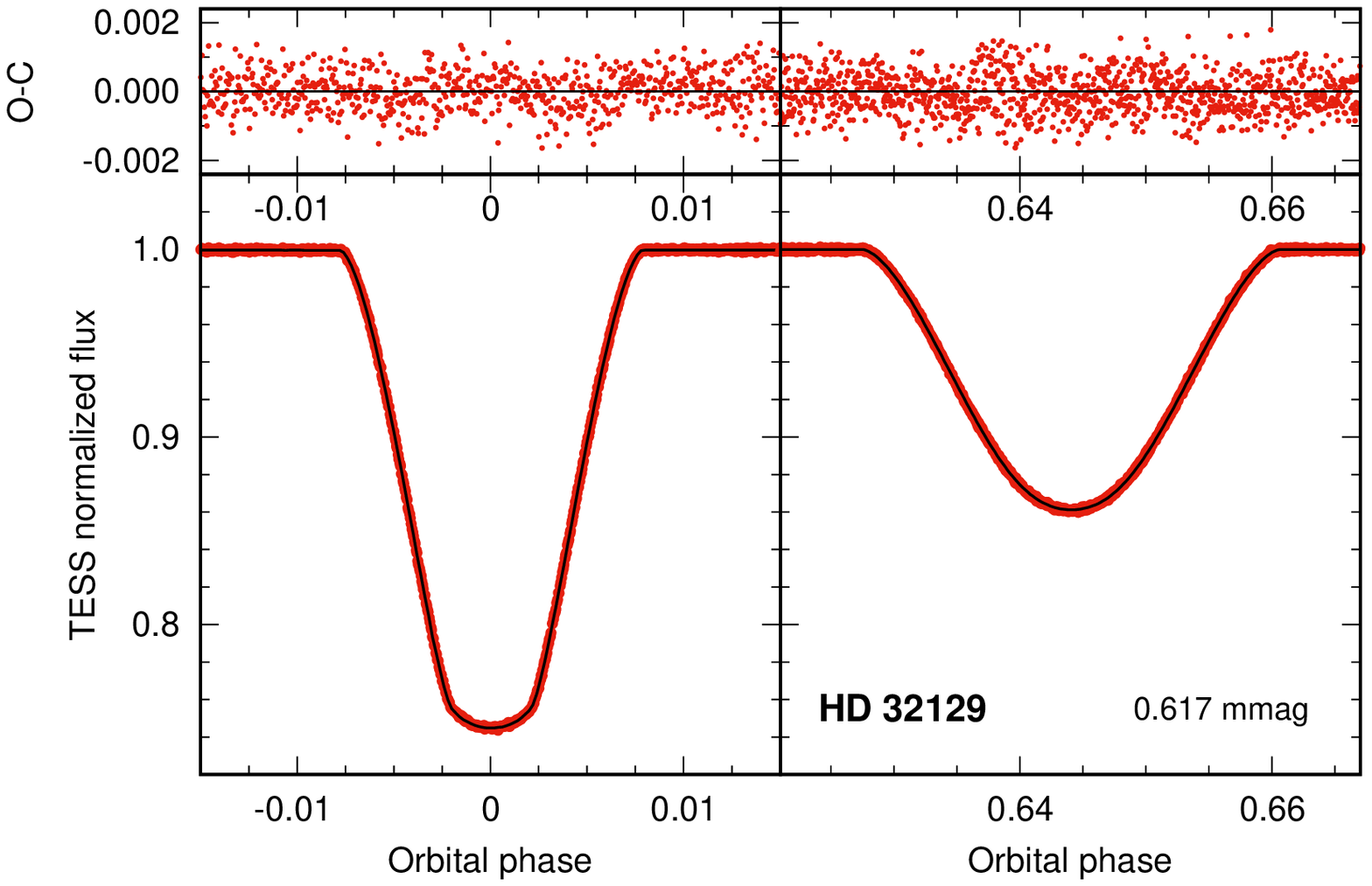}\vspace{-1.45cm}
\mbox{}
\includegraphics[angle=0,scale=.46]{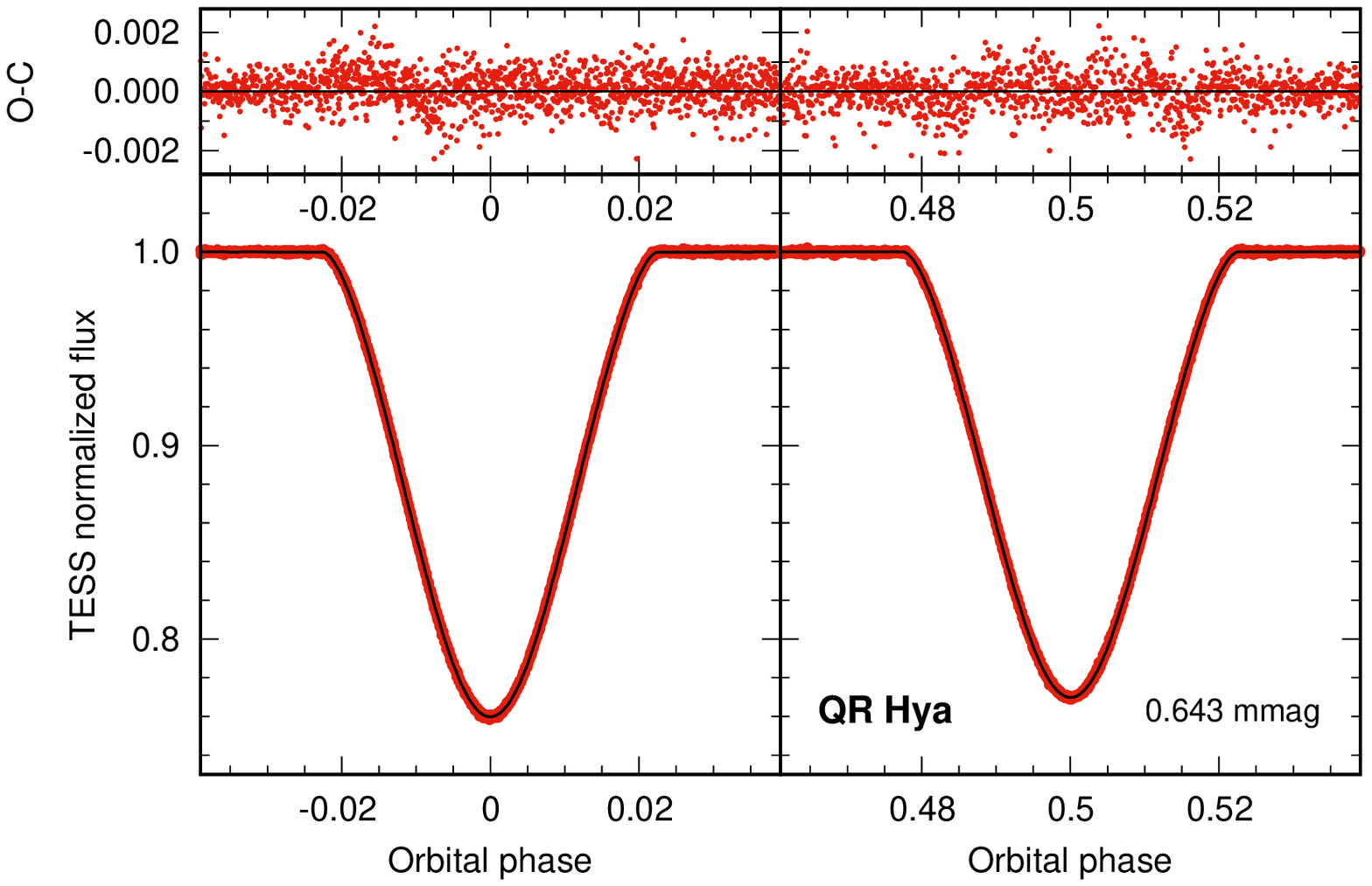}\vspace{-1.45cm}
\mbox{}
\includegraphics[angle=0,scale=.46]{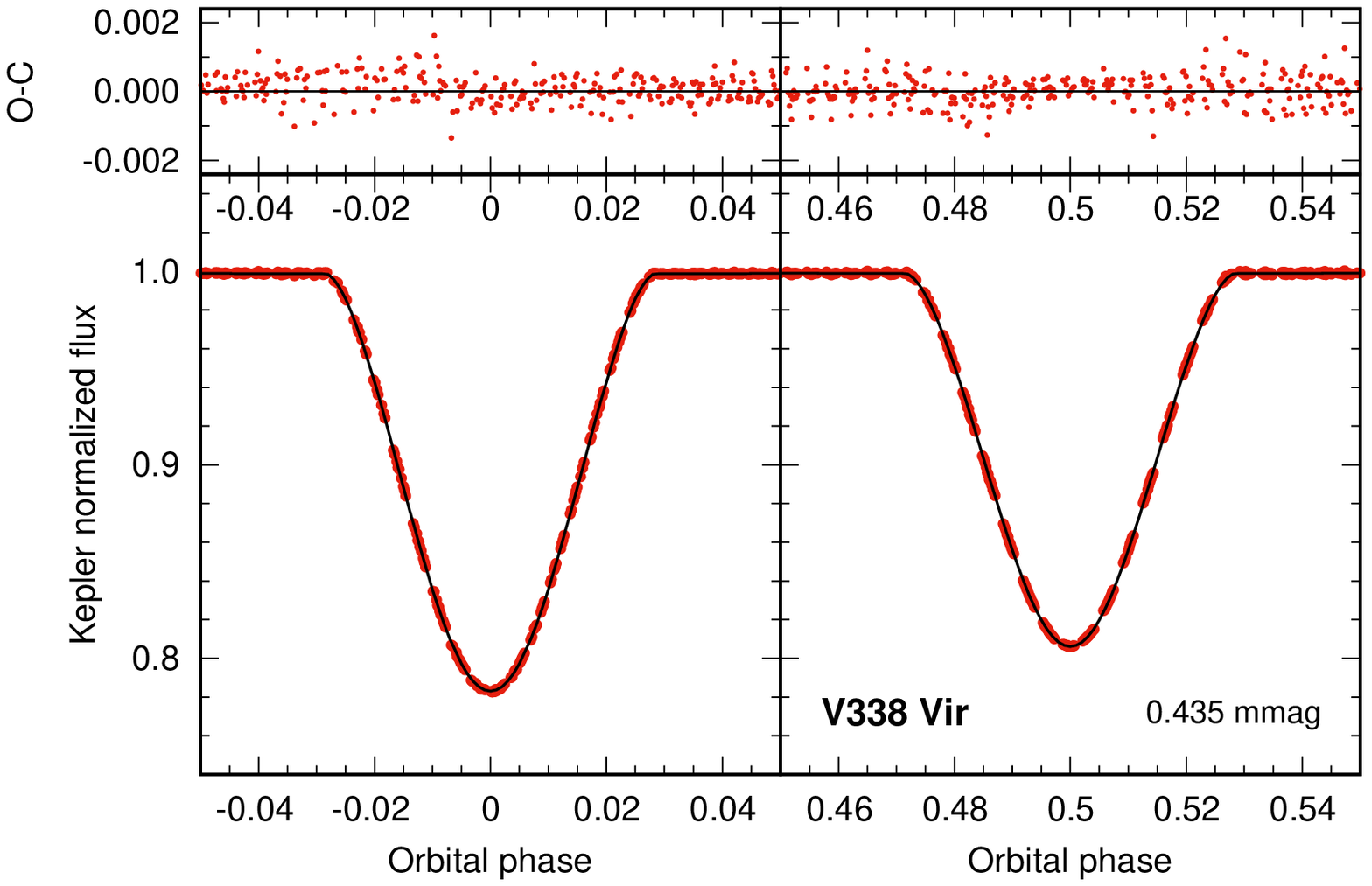}\vspace{-1.45cm}
\mbox{}
\includegraphics[angle=0,scale=.46]{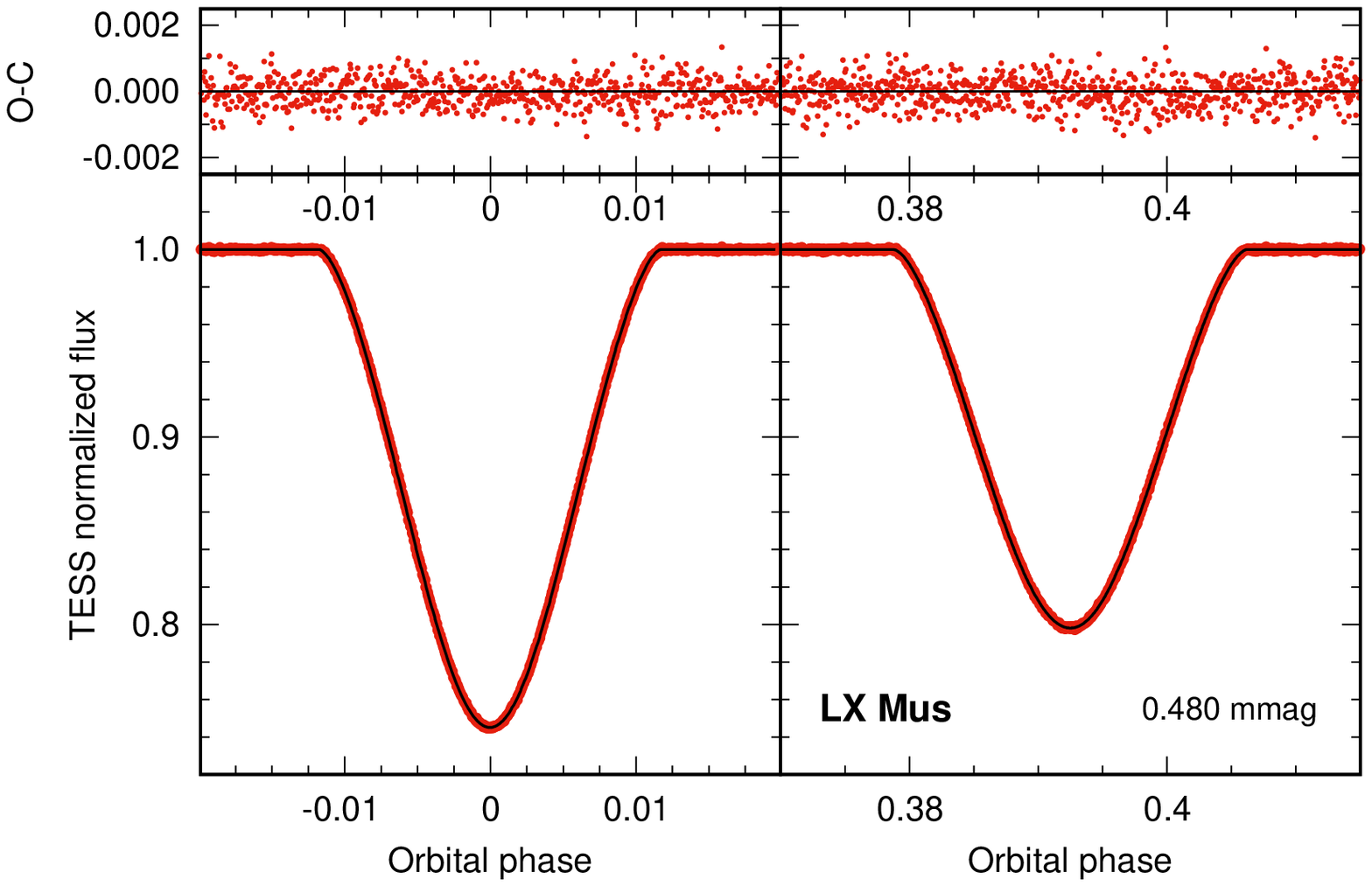}\vspace{-1.45cm}
\mbox{}
\includegraphics[angle=0,scale=.46]{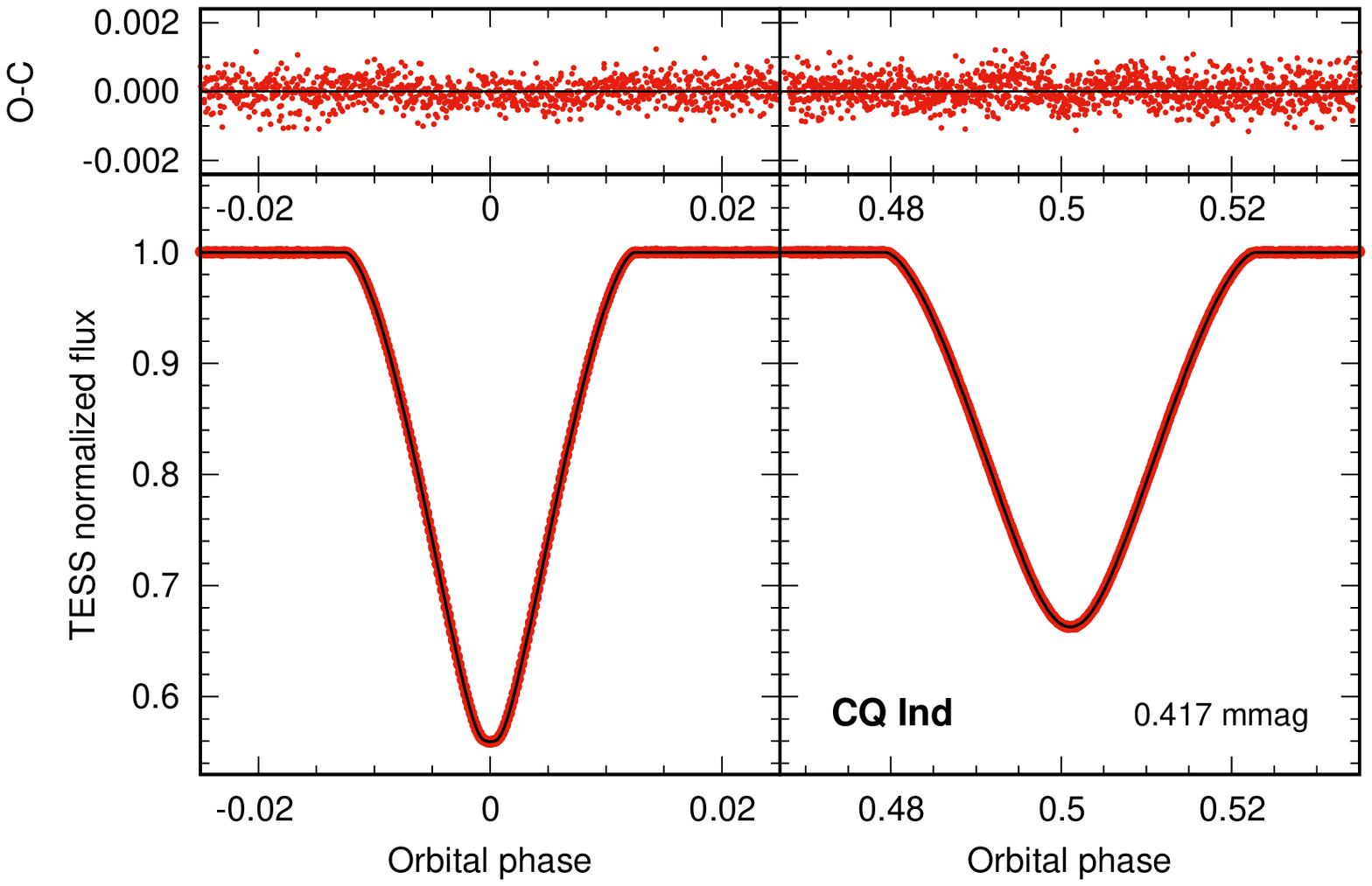}
\end{minipage}\hfill
\caption{The WD model fits to the photometric observations. Red points - observations, black line - synthetic light curve. The {\it rms} of the best solution is given in the lower right corner. \label{fig:light}}
\end{figure*}

\begin{figure*}
\begin{minipage}[th]{0.5\linewidth}
\vspace{-1.1cm}
\includegraphics[angle=0,scale=.43]{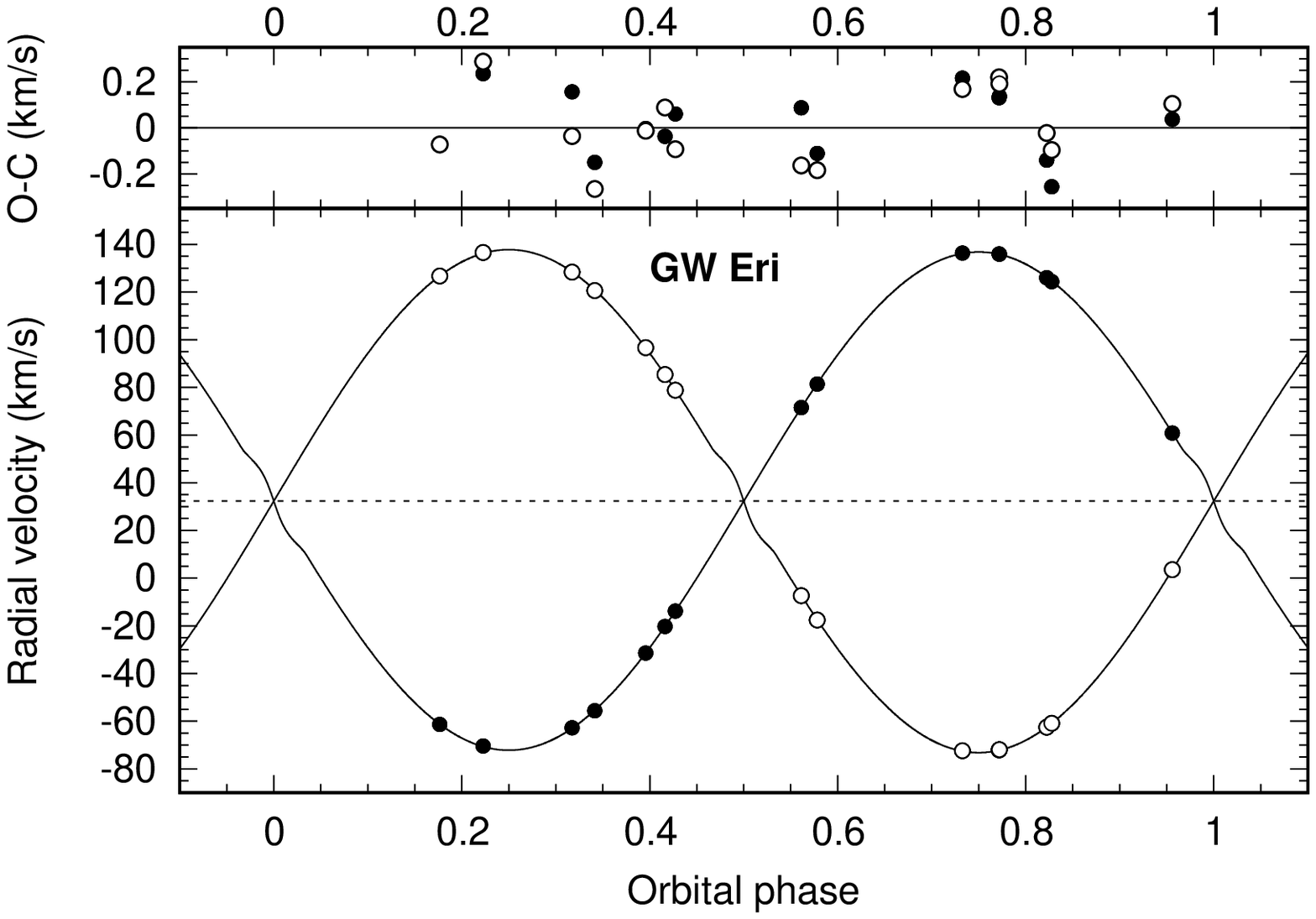} \vspace{-1.cm}
\mbox{}
\includegraphics[angle=0,scale=.43]{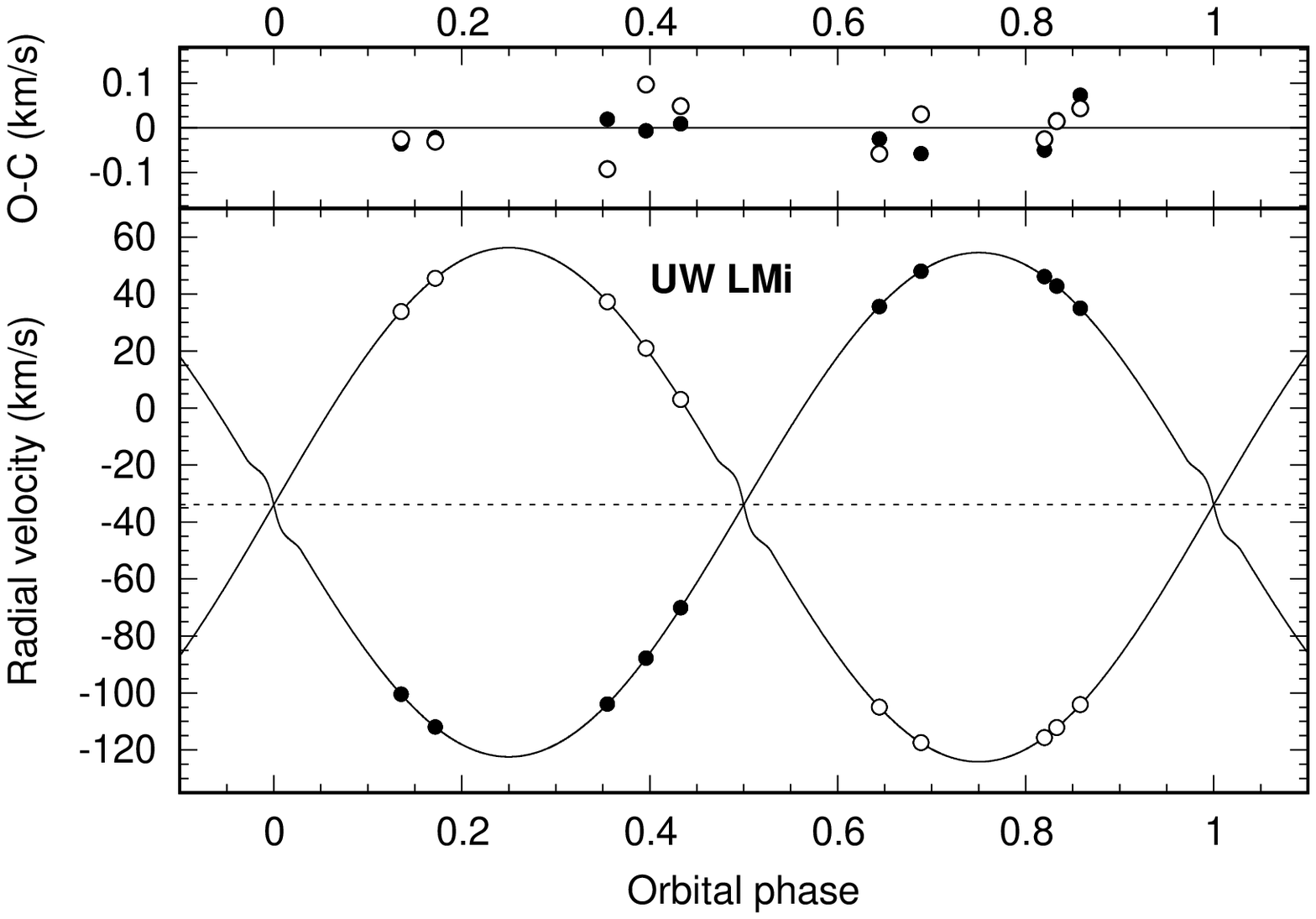} \vspace{-1.cm}
\mbox{}
\includegraphics[angle=0,scale=.43]{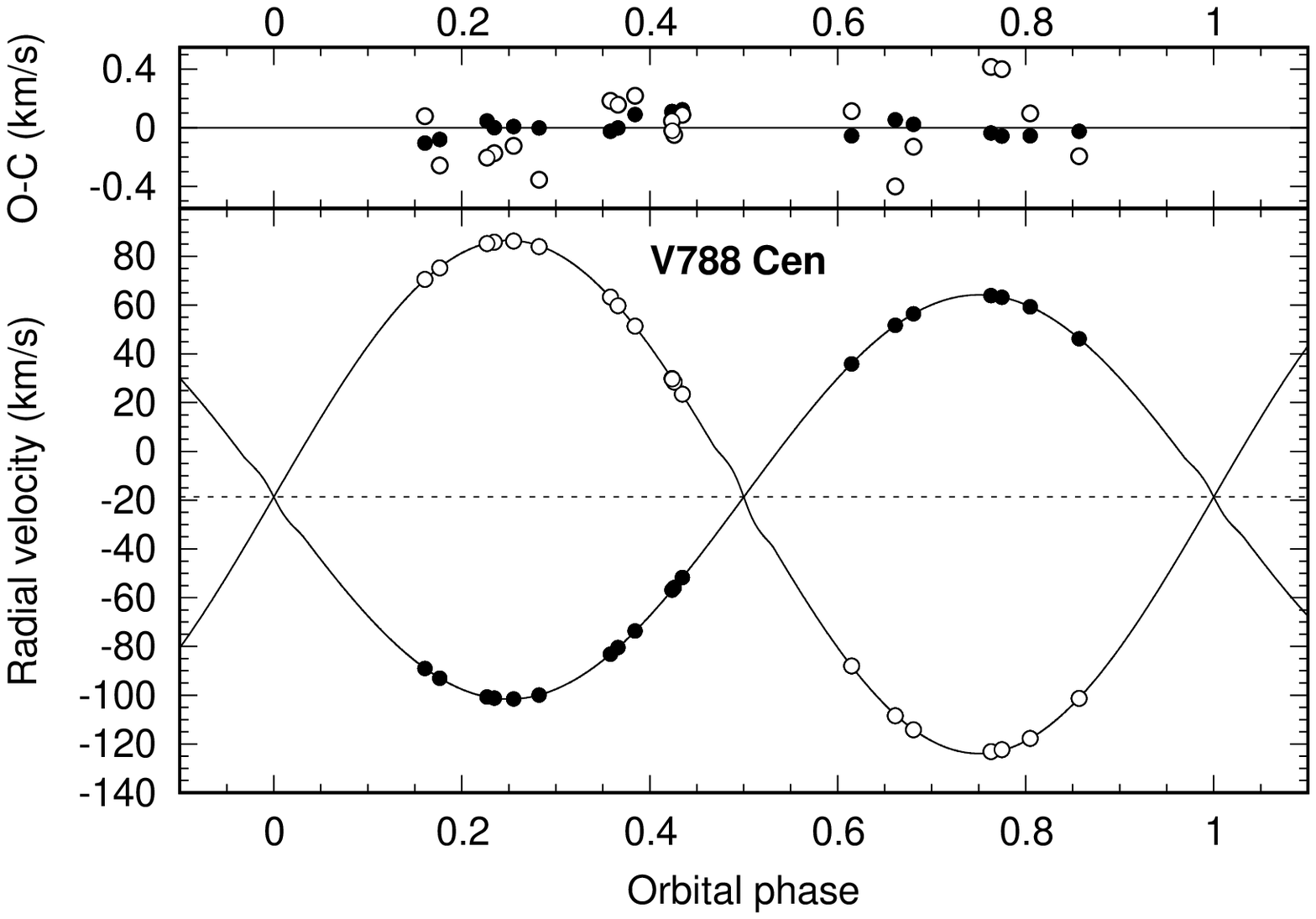}\vspace{-1.cm}
\mbox{}
\includegraphics[angle=0,scale=.43]{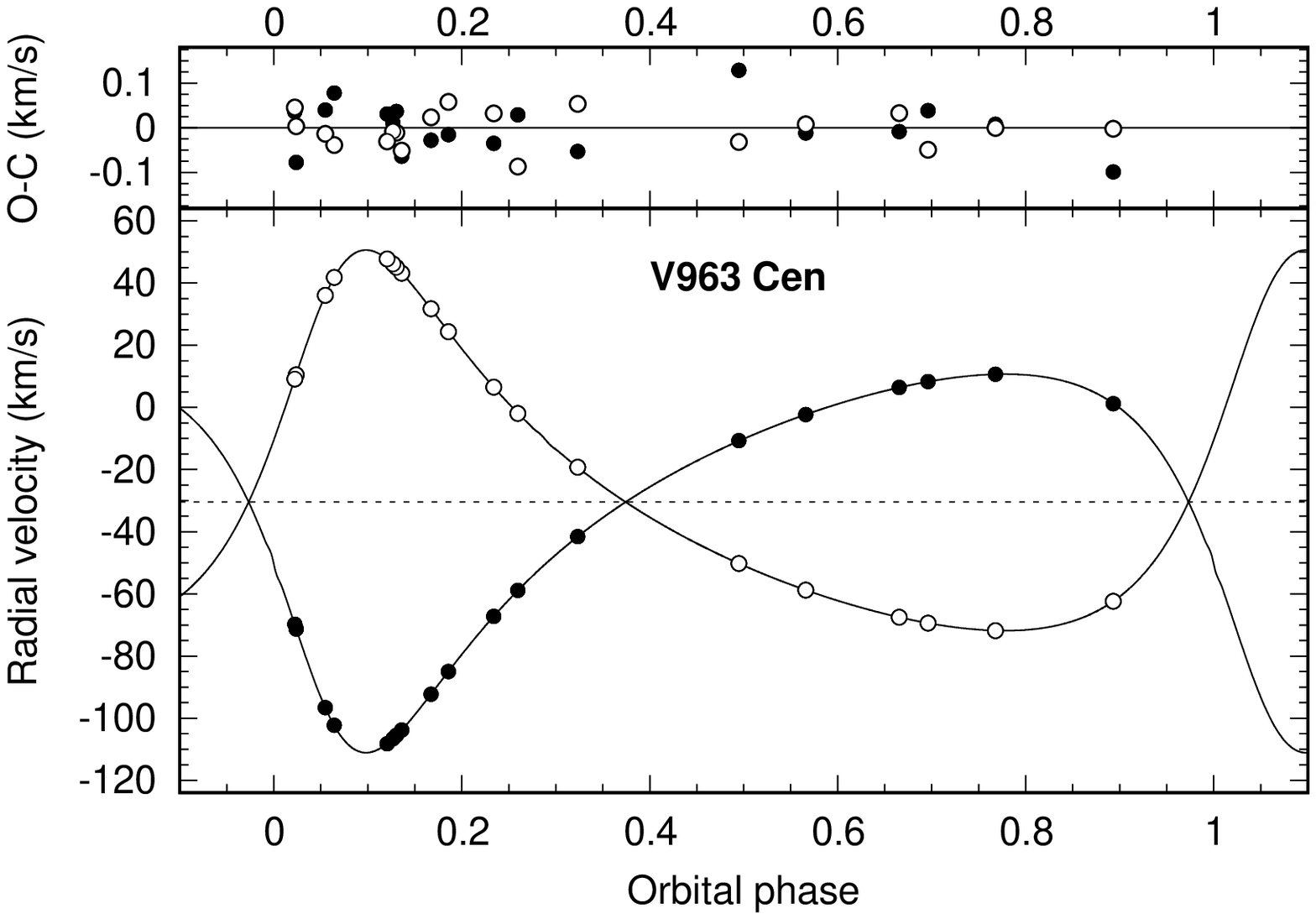}\vspace{-1.cm}
\mbox{}
\includegraphics[angle=0,scale=.43]{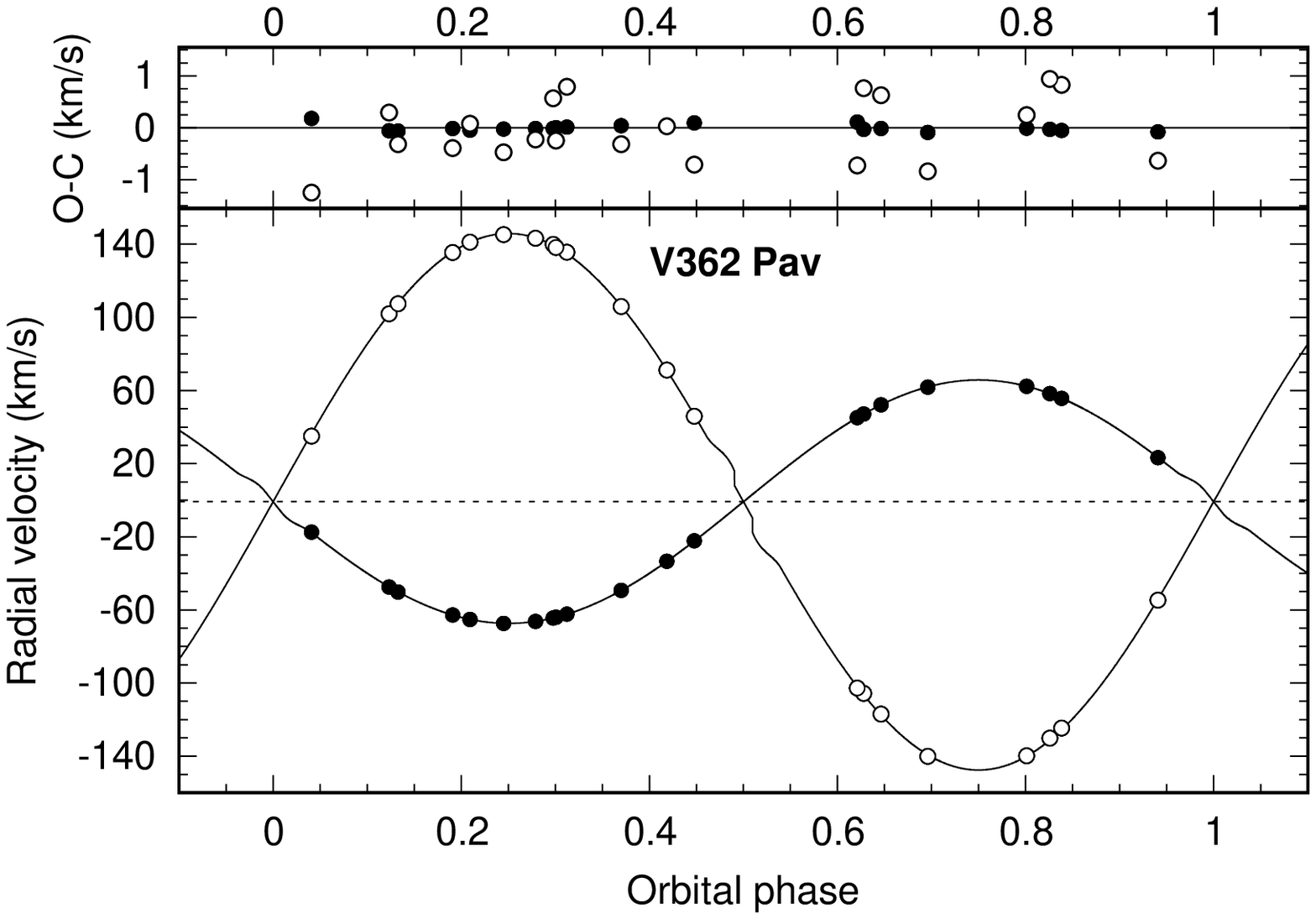}
\end{minipage}\hfill
\begin{minipage}[th]{0.5\linewidth}
\vspace{-1.1cm}
\includegraphics[angle=0,scale=.43]{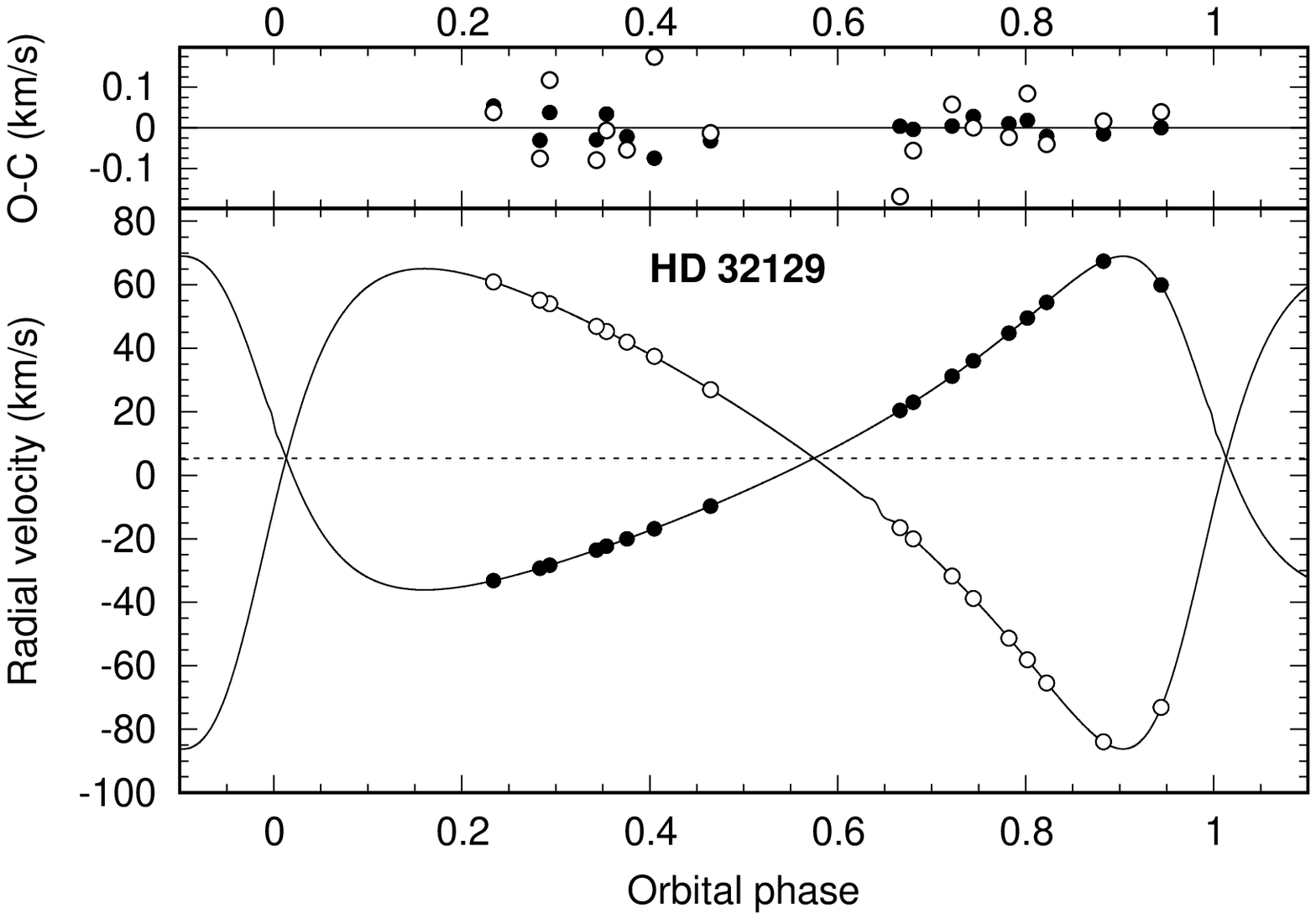}\vspace{-1.cm}
\mbox{}
\includegraphics[angle=0,scale=.43]{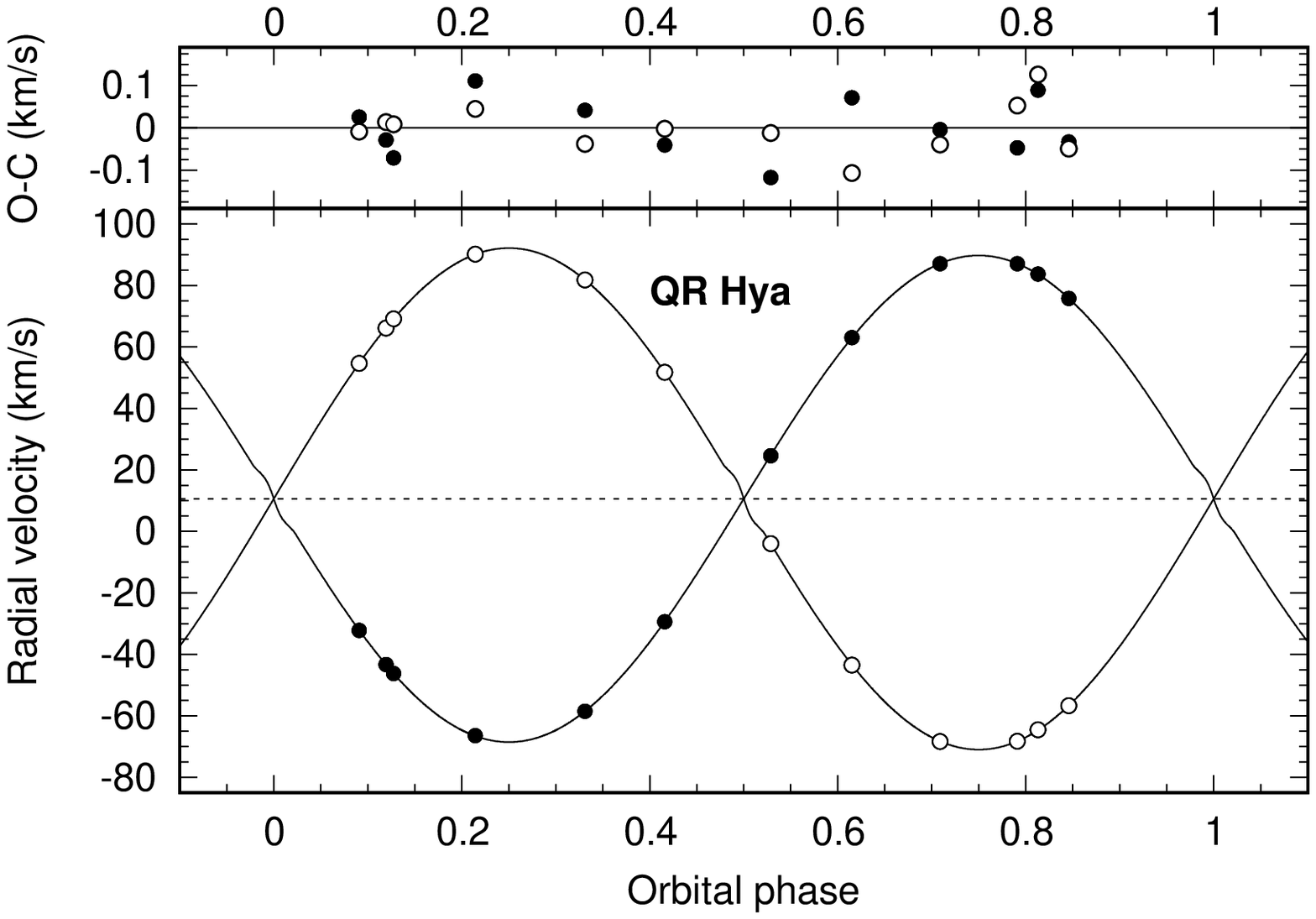}\vspace{-1.cm}
\mbox{}
\includegraphics[angle=0,scale=.43]{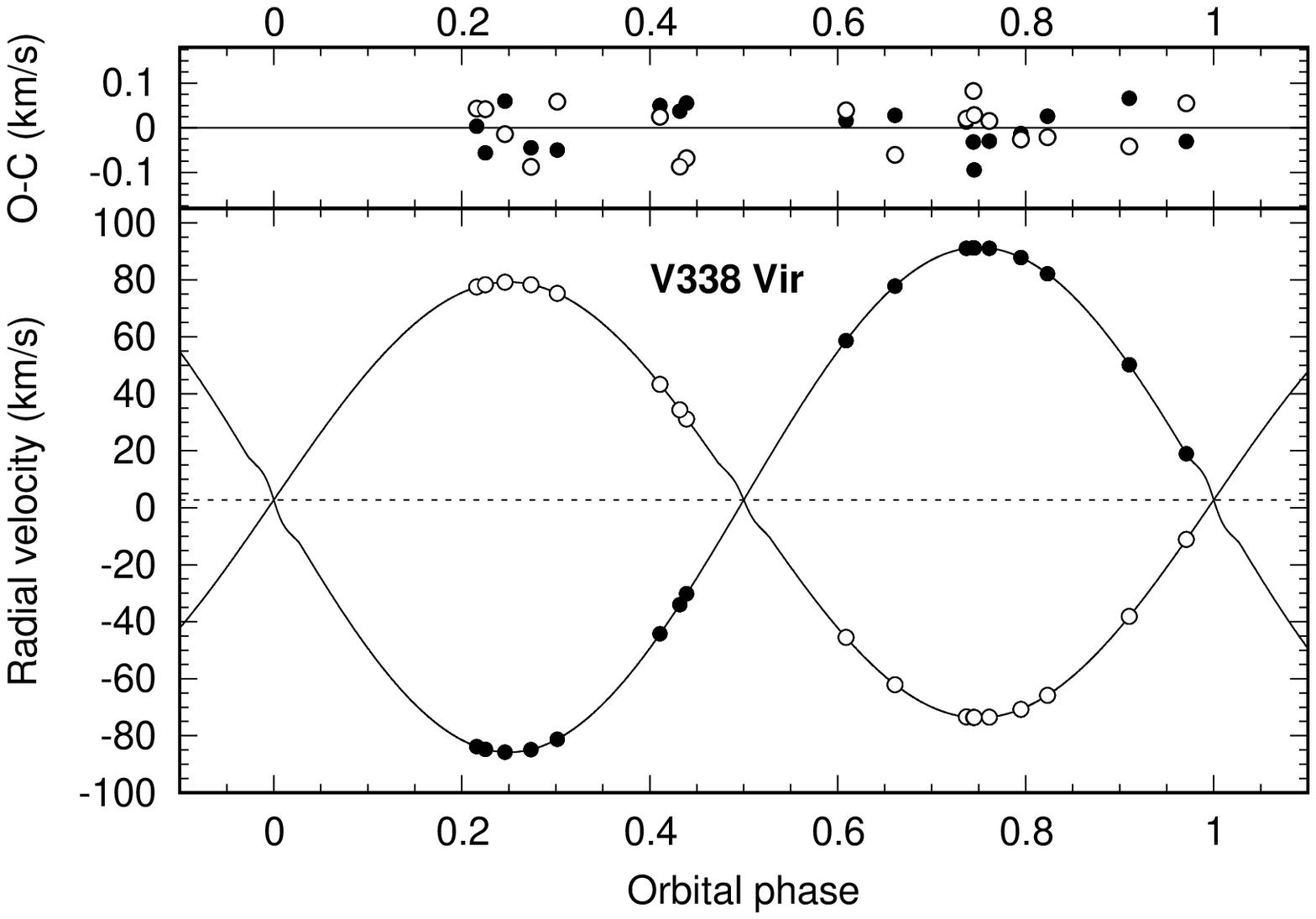}\vspace{-1.cm}
\mbox{}
\includegraphics[angle=0,scale=.43]{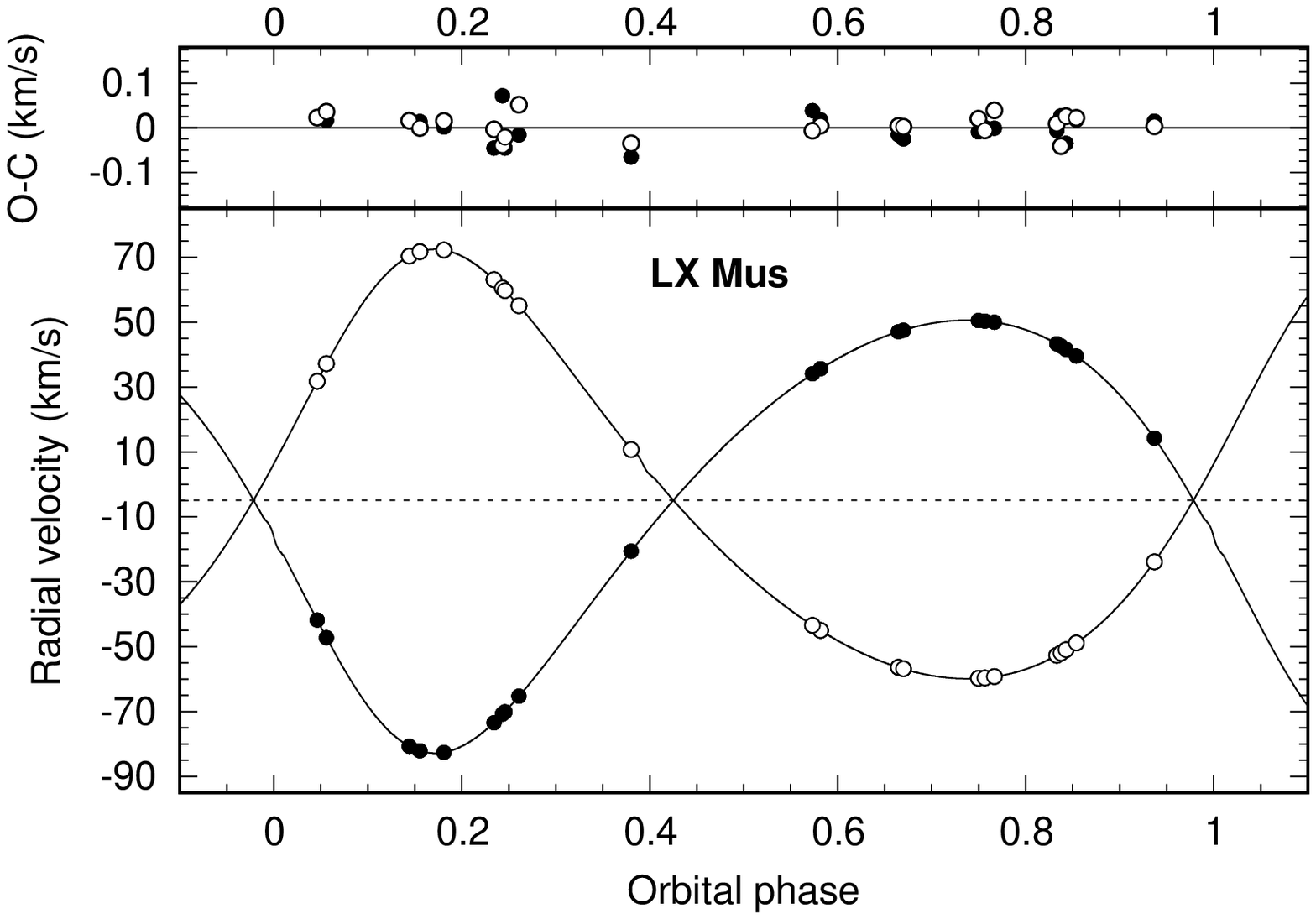}\vspace{-1.cm}
\mbox{}
\includegraphics[angle=0,scale=.43]{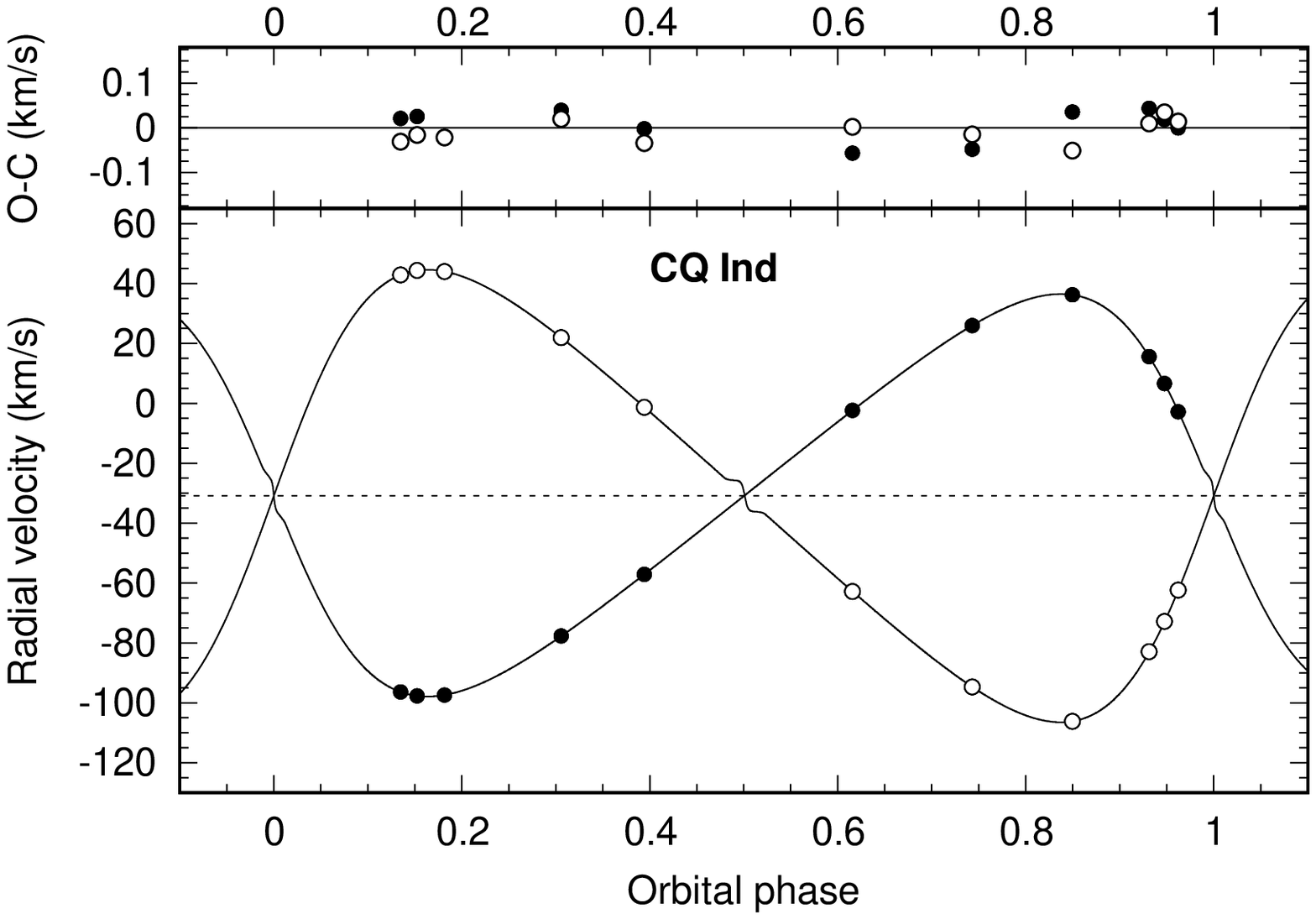}
\end{minipage}\hfill
\caption{The WD model fits to the HARPS velocimetry: filled circles - the primary, open circles - the secondary. Black lines - model predictions. \label{fig:rv}}
\end{figure*}

The statistical (formal) errors on the fitted parameters were estimated with the Differential Correction subroutine of the WD code. We assumed very conservative errors on parameters: we multiplied the formal errors by a factor of 3. The model synthetic light curves compared against photometric observations are presented in Fig.~\ref{fig:light} for all ten systems with the rms of the best solution given. The radial velocity solutions plotted against observed velocimetry are presented in Fig.~\ref{fig:rv}.

The model parameters for all systems are summarized in Table~\ref{tab_par_orb}. The systemic velocity is not corrected for the gravitational redshift or convective blueshift. The absolute dimensions of the systems were calculated using nominal astrophysical constants advocated by IAU 2015 Resolution B3 \citep{prsa16} and are presented in Table~\ref{par_fi}.

\subsection{Analysis details and results}
\label{WD_results}
\subsubsection{GW~Eri}
This is the most massive and hottest eclipsing binary in our sample, and is a triple system. It forms a common proper motion pair with HD 26590 which lies at a distance of about 60 arcsec. The TESS light curve shows two partial eclipses of moderate and similar  depth. The orbit is circular and the components are similar in their physical properties. The similarity of the components and partial eclipses leads to a strong correlation between their radii. In order to break it we determined a $V$-band light ratio from available HARPS spectra. Strangely, the spectroscopic light ratio is significantly different during both quadratures: spectra taken between orbital phases 0.15 and 0.45 show the secondary being consistently brighter than the primary by about 3\% while all spectra taken between orbital phases 0.55 and 0.97 show the opposite effect with the primary being 5\% brigher. The reason for this difference is unclear and it may be connected with the metallic nature of both components. In the subsequent analysis we did not constrain our models for the light ratio. The final solution needs a small amount of third light in order to remove systematic residuals in both eclipses. No very close optical companions to GW Eri is known. It is possible that the detected third light is stray light from the nearby HD 26590, which is only 2 mag fainter than the system.

Both stars rotate synchronously and the estimated spectral types of the components are A4m\,V+A4m\,V but due to the strongly metallic nature of both components they are somewhat uncertain: both stars are hotter than expected for their masses. There is a significant spread of spectral type classifications in the literature: \cite{hou88} give A1mA2-A8,  \cite{abt95} give kA2hA5VmF2 while \cite{AbL77} give kA1hA3VmA3 + kA1hA3VmA3. Although the spectra of both components are rich in metallic lines, the relatively fast rotation means that radial velocity measurements are only precise to about 140~m~s$^{-1}$. The $rms$ of the radial velocity solutions are 137 and 154 m~s$^{-1}$, so are fully consistent with measurement errors. We do not detect any radial velocity trends during the timespan of 12 years covered by the HARPS spectra.

\subsubsection{HD~32129}
This well-detached system shows a distinct total primary eclipse and a shallow partial secondary eclipse, due to a significant orbital eccentricity ($e\sim 0.44$) and a high orbital inclination ($i\sim89$ deg). The relatively long phase of totality during primary eclipse suggests large difference in size between the two components. The out-of-eclipse parts of the TESS and \textit{Kepler} light curves are flat, with a tiny flux modulation probably due to small starspots. The \textit{Kepler} K2 long-cadence light curve was rectified using the method used for FM~Leo in our previous work \citep{gra21}. We tried to solve the TESS and K2 light curves simultanouesly with the radial velocity curves. However, because of long time interval covered by both types of data (the spectra were taken over $\sim 1350$ day and the K2 and TESS epochs are separated by $\sim 1600$ days), apsidal motion has a significant effect on the times of eclipse and the shape of radial velocity curves. We therefore initially solved the light curves and radial velocities separately. After several iterations we could find common orbital parameters (eccentricity $e$, longitude of periastron $\omega$ and rate of periastron advance $d\omega/dt$) and thus carry out a full simultaneous solution. The apsidal motion has a rate of $1.9\cdot10^{-4}$ deg cycle$^{-1}$, which corresponds to an apsidal motion period of $\sim$85\,000 years.

In order to properly fit the shape of the primary eclipse we had to adjust also the third light in both filters (K2, TESS). The calculated third light contributes about 1\% of the flux in the K2 band and 2\% in the TESS band, so is redder than the light from the eclipsing system. If we assume that it comes from a physically bound companion to the system, i.e.\ at a common distance, it would correspond to absolute magnitudes of $M_{Rc}\!=\!7.4$ mag and $M_{Ic}\!=\!6.2$ mag \footnote{For HD 32129 the observed magnitudes are $R_C=8.77$ mag (K2), $I_C=8.49$ mag (TESS) and reddening $E(B-V)=0.11$ mag.}. Both numbers are consistent with a K6/K7 dwarf. Its expected contribution to the total light in the $K$-band is 5.5\%. The final fit to TESS light curve (Fig.~\ref{fig:light}) is very good in the case of the secondary eclipse but there are small systematic deviations (up to 300~ppm) in the primary eclipse near fourth contact. The radial velocity fit (Fig.~\ref{fig:rv}) is fully acceptable with minor residuals for the primary (the $rms$ is only 30~m~s$^{-1}$) and significantly larger for the secondary (78~m~s$^{-1}$). The secondary's larger $rms$ is a result of it being significantly fainter than the primary: in the $V$-band it is 7 times fainter than the primary.

The primary star is much more massive, hotter and larger than the companion. The secondary is a solar-twin star regarding its size, mass and temperature, however its surface composition is more metal rich. Both components rotate very slowly ($v\sin{i}\sim$~1--2~km~s$^{-1}$). The estimated spectral type is F3\,V + G2\,V.

\subsubsection{UW~LMi}
\label{uwlmi}
This is the only system in the sample which has no light curve based on space photometry. Instead we used ground-based Str{\"o}mgren photometry in the $uvby$ bands. However its precision is almost an order of magnitude lower than photometry from TESS or K2 for targets of similar brightness. Fortunately the use of four different bands mitigates this effect, in particular allowing a secure determination of the temperature ratio of the components. The system shows two relatively deep, partial eclipses of similar depth. The orbit appears to be circular and the out-of-eclipse parts of the light curves are practically flat. Because of the lower precision of the light curves we exceptionally used less dense grids on stellar surfaces for the WD analysis, with \verb"N1=N2"=50.

In order to improve the solution we applied a spectroscopic light ratio in the $V$-band as an additional constraint. It is well determined from HARPS spectra: $L_2/L_1=0.882\pm0.008$. We carried out a simultanenous solution of the $uvby$ data and radial velocity curves. The calculations converged to a solution with the primary being slightly more massive, hotter and larger than the companion. The light ratio in the $V$-band predicted by the model is 0.880 so is in perfect agreement with the spectroscopic value. We initially included third light as a fitted parameter, but the WD code returns always small negative values with no improvement in the residuals, so we set $l_3=0$ in the main analysis. The final solution in the $y$-band is presented in Fig.~\ref{fig:light} and shows a good fit to the observations, free of systematic deviations. The $rms$ is gradually decreasing with wavelength from $7.9$~mmag in $u$ to $5.5$~mmag in $y$. The solutions of the radial velocity curves are of high quality: the $rms$ for the primary is 37~m~s$^{-1}$ and for the secondary is 54~m~s$^{-1}$. The rotation of both components is fully synchronized with the orbital period: $v\sin{i}\approx v_{\rm syn}\approx 17$~km~s$^{-1}$. We do not find evidence for significant period changes in the system: we find consistent parameters of $P_{\rm orb}$ and $T_0$ in the simultaneous solution and from radial velocities solved separately, despite the mean epoch of the velocimetry being 7000 ~d ($\sim$1800 orbital cycles) later than the epoch of the photometry.

The estimated spectral type is G0\,V + G0\,V. This is fully consistent with the classifications of G0\,V given by \cite{upg70} and G0\,V + G1\,V given by \cite{Gri01}.

\subsubsection{QR~Hya}
A WD model of the system was obtained by fitting the TESS light curve from sector 9 and the HARPS velocimetry. The system shows partial eclipses of almost equal depth, and the secondary minimum is at orbital phase 0.5. We first fitted a model with a circular orbit and no third light. The orbital period was assumed to be constant. The iterations easily converged to a solution where the primary star is the slightly hotter, larger and more massive component. This solution gives a $V$-band light ratio of $L_2/L_1=0.82$ which is fully consistent with the spectroscopic light ratio of $0.80\pm0.02$. However, some systematic residuals of up to 1000 ppm versus the best fit remained in both eclipses. Including the orbital eccentricity and the argument of periastron as free parameters allowed a significantly better solution to be obtained, both for the light and radial velocity curves, although the resulting eccentricity is very small ($e\sim0.0001$). Inclusion of third light as a free parameter does not improve the solution and the WD code always returns small negative values of $l_3$. The final solution still shows some small systematic residuals (see Fig.~\ref{fig:light}) during eclipses but they are likely an artefact of removing trends in the TESS light curve caused by spot activity.

The residuals of the radial velocity solution show an $rms$ of about 60~m~s$^{-1}$ for both components. They are consistent with the precision of the radial velocity determination: the typical S/N of the spectra is not high (Table~\ref{tab:harps}) and the BF profiles are slightly rotationally broadened ($v\sin{i}\approx13$~km~s$^{-1}$). The rotation is fully synchronous and the tidal deformation of the components, defined as $(r_{\rm point}-r_{\rm pole})/r_{\rm mean}$, is just 0.1\%. Both components are slightly-evolved solar-type stars with a practically solar metallicity of [M/H] $=-0.01$ dex.

We estimated the spectral type of the system as G1\,V + G2\,V based on the calibration by \cite{pec13}. This is fully consistent with the spectral type G1\,V reported by \cite{hou82} based on photographic plates as well with the G1\,V + G2\,V based on high-resolution spectroscopy \citep{cut02}.

\subsubsection{V788 Cen}
\label{v788cen}
The TESS light curves from sectors 10 and 37 were combined with the radial velocity curves in order to obtain a simultaneous solution with the WD code. The light curves show two shallow, partial eclipses of slightly unequal depth. The orbital phase of the secondary minimum is exactly 0.5 and the ephemeris given by \cite{Cou74} accurately predicts the eclipse times in the TESS data. We assumed a circular orbit and constant orbital period during the first stage of our analysis. The iterations with WD quickly converged to a solution with the hotter and more massive primary being almost twice as large as the secondary. The resulting light ratio in the TESS passband, $L_2/L_1=0.3$, corresponds to a $V$-band light ratio of about 0.29, in good agreement with the observed intensity of absorption lines in the spectra ($\sim$0.30). The model we obtained has a small but non-negligible third light, $l_3\sim0.02$.

Futher investigation of residuals revealed abrupt changes in flux in the TESS light curves of instrumental origin, reaching 0.05\% of the total flux, and also small flux trends lasting up to few days due to spot activity and/or slow instrumental drifts. We corrected for them and repeated the fitting procedure, obtaining significant decrease of residuals in both eclipses. However, the third light was persistent. If this light would come from a physically bound close tertiary component it would have $M_{Ic}\!=\!4.9$ mag which would correspond to a K0\,V spectral type. The typical precision of individual radial velocity determination is $\sim 65$~m~s$^{-1}$ for the primary and $\sim 180$~m~s$^{-1}$ for the secondary. This precision is in accordance with the $rms$ of the primary's residuals (Fig.~\ref{fig:rv}) but the secondary's residuals are somewhat large. This may suggest some non-radial pulsations on the surface of the secondary star.

The temperature difference $T_1-T_2$ inferred from light curve solution is $414\pm30$ K which is fully consistent with the temperature difference derived from spectroscopy (Section~\ref{temp:atm}) $360\pm110$ K. The components differ in mass and radius, with the primary being a significantly more evolved star. The metallicity is super-solar and the primary rotates slower than synchronous. We estimated the spectral type as A7\,IV + A9\,V, which is somewhat later than the types reported before: A3m by \cite{And77} and A2mA5-F2 by \cite{hou78}.

\subsubsection{V338 Vir}
The components of this system differ significantly in size, but their temperaures are only slightly different. The more massive star is cooler and is eclipsed during the secondary minimum. The orbit is circular and the eclipses are of moderate depth. The out-of-eclipse part of the K2 light curve is not flat but shows a small ellipsoidal effect with an amplitude of $\sim$0.002~mag. The K2 long-cadence light curve was rectified using the method applied to FM Leo in our previous work \citep{gra21}. We carried out a simultaneous solution of photometry and velocimetry, and as the secondary is the more luminous star we adjusted temperature $T_1$ instead of $T_2$. Because the eclipses are partial and shallow we included the spectroscopic light ratio as an additional constraint. From a number of HARPS spectra we determined a $V$-band light ratio of $L_2/L_1=2.15\pm0.05$ and we forced the WD solutions to reproduce this light ratio to within its error bars.

The final light curve solution is presented in Fig.~\ref{fig:light}. It shows a number of instrumental trends and effects, which were only partially removed during the detrending procedure in order to preserve the out-of-eclipse proximity effects. The overall fit is very good, especially for the primary minimum, and only in the secondary minimum small systematic deviations of up to 200 ppm are present. Inclusion of third light does not improve the fit. The $rms$ of the fits to the radial velocities are 46~m~s$^{-1}$ and 51~m~s$^{-1}$, and they are consinstent with the precision of individual radial velocity measurements. Both stars rotate sub-synchronously and their metallicity is sub-solar. The secondary is an evolved star and it is close to a subgiant phase. The estimated spectral type of the system is F5\,V + F6\,V-IV, in agreement with the classification of F5\,V given by \cite{hou99}.

\subsubsection{V963 Cen}
\label{v963cen}
The system shows a relatively deep primary eclipse and a shallow secondary eclipse, both partial. The orbit is significantly eccentric ($e\sim0.42$). To obtain a WD model of the system we combined TESS photometry with the Str{\"o}mgren $uvby$ photometry and HARPS velocimetry. It turned out that obtaining a fully consistent simultaneous solution was very difficult and practically impossible for a number of reasons. First, the system shows apsidal motion with a period of about 55\,000~yr, complicating analysis of data obtained over a long time interval. The mean epochs of the observations are JD 2451000, 2456400 and 2459350 for the $uvby$ photometry, velocimetry and TESS photometry, respectively. Second, the analysis based on velocimetry or TESS photometry leads to different orbital eccentricities: the photometric one is 0.4237, whilst the spectroscopic one is 0.4218; the difference is more that 6$\sigma$. Third, the analysis of TESS photometry alone leads to temperature ratio $T_2/T_1=0.990$ while analysis of disentangled spectra, as well as $uvby$ photometry solved alone, gives $T_2/T_1>1$. Fourth, a solution derived from $uvby$ photometry gives a significantly larger orbital inclination than one based on the TESS photometry.

In order to find a consistent solution we fitted separately the three blocks of data, with the aim of obtaining as many consistent orbital and photometric parameters as possible. Full agreement was found for $P_{\rm orb}$, $\Omega_1$, $\Omega_2$ and $e$. The third and fourth problems were much mitigated by adjusting the third light: it turned out that TESS light curve has a quite large negative $l_3$, assuming that $uvby$ photometry has zero third light. The second problem was solved by finding a compromise eccentricity of 0.4223, while the first problem was overcomed by adjusting $\omega$ and $\phi$ separately in each block of data assuming the same eccentricity. The solution of the TESS light curve is presented in Fig.~\ref{fig:light}. The fit shows small but noticable systematic deviations during secondary eclipse -- those residuals can be removed by increasing the eccentricity but at a cost of degrading the radial velocity solution. The fit to the velocimetry is presented in Fig.~\ref{fig:rv}. The $rms$ of the residuals is 55~m~s$^{-1}$ for the primary and 37~m~s$^{-1}$ for the secondary.

The components of the system are very similar to each other: they differ in surface temperature by only $\sim$10~K, in mass by 0.5\%, and in radius by 1.7\%. The primary is the more massive, larger and cooler of the two. The components rotate about two times faster than synchronous rotation, but two times slower than the synchronous value at periastron. The estimated spectral type is G2\,V-IV + G2\,V-IV, which corresponds well with the classification of G2\,V given by \cite{hou75}.

\subsubsection{LX Mus}
The system consists of two very similar stars on an eccentric orbit. The eclipses are partial and rather shallow, and the out-of-eclipse parts of TESS light curve are flat with only tiny modulations due to some small stellar spots. The simultaneous solution of the photometry and velocimetry quickly converged to a solution with a slightly less massive, cooler and smaller star eclipsed during the deeper primary minimum. We also included third light in the fit, but adjusting this parameter does not reduce the residuals so we subsequently assumed $l_3=0$. The apsidal motion is not detected with the observations used in our analysis: the data likely cover too short a time interval to do so. We also do not detect any orbital period changes. The predicted light ratio at the $V$-band is $L_2/L_1=1.11$, which is slightly inconsistent with the spectroscopic light ratio of $1.06\pm0.02$. Forcing the WD code to reproduce the value of the spectroscopic light ratio worsens the fit and produces some small but noticable systematic deviations in both eclipses. To take this inconsistency into account we enlarged the errors on the radii by a factor of 1.5.

The $rms$ of radial velocity residuals are very small: 28~m~s$^{-1}$ and 25~m~s$^{-1}$ for the primary and the secondary, respectively. Those values are consistent with the precision of the radial velocity determinations: the numerous and sharp lines allow for a very precise determination of the broadening function profile.  Both components rotate slightly slower than synchronous. The estimated spectral type of the system is F5\,V + F5\,V, which is in perfect agreement with the F5\,V reported by \cite{hou75}.

\subsubsection{V362 Pav}
The system consists of two very different stars on a practically circular and relatively tight orbit. The light ratio in the $V$-band is 70 and the primary completely dominates the spectrum. Fortunately the system has a favourable geometry with total eclipses, which allows the radii of both components to be precisely determined. The TESS light curve shows a noticable ellipsoidal effect with an amplitude of $\sim$0.01~mag. The secondary eclipse is much shallower than the primary eclipse, indicating a large temperature difference between the components. Radial velocity curves show a large difference in the component's masses, with a mass ratio of $q\sim0.45$.

Simultaneously fitting the photometry and velocimetry proved to be difficult. Although the fits converged quickly to a solution, inspection of the light curve residuals revealed that the model gave large systematic effects during eclipses (especially the primary) and a sinusoidal-like pattern of residuals outside of eclipses. We included third light as an adjusted parameter and found that it improved the fit but at the expense of a significantly negative value: $l_3=-0.054\pm0.005$. The systematic residuals during the secondary eclipse almost vanished but remained (albeit much diminished) in the primary eclipse. We decided to adjust also the albedo parameters for both components, finding that the albedo of the primary was very low and consistent with zero whilst that of the secondary was close to the value of 0.5 expected for a star with a convective atmosphere. However, this did not completely get rid of the systematic residuals especially close to the second and third contacts in the primary eclipse.

\begin{figure}
\hspace*{-0.5cm}
\includegraphics[angle=0,scale=0.5]{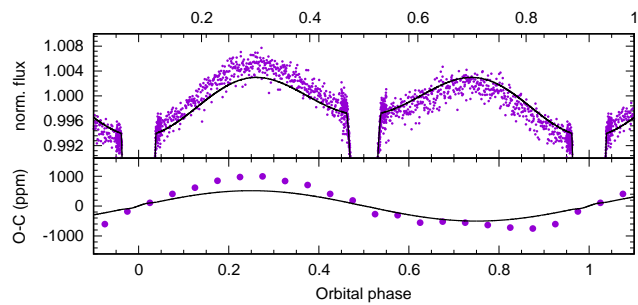}
\caption{{\it Top}: out-of-eclipse flux changes of V362~Pav, with the best fit model from WD overplotted. The ellipsoidal modulation and the reflection effects are included in the model. {\it Bottom}: residuals binned in steps of 0.05 in phase. The theoretical light variation due to Doppler beaming is overplotted.\label{DopBin}}
\end{figure}

Then we turned out attention to the out-of-eclipse residuals. The strong difference in brigthness between components and relatively high radial velocity semiamplitude of the primary suggested that Doppler beaming will be significant in this system. Fig.~\ref{DopBin} shows the out-of-eclipse variations of V362 Pav together with a model fit (upper panel). The lower panel shows binned residuals from the model plotted against the theoretical Doppler beaming effect (a line). For calculating the effect we used equation 9 from \cite{plac19}. We estimated the beaming factor for the primary, assuming $T_{\rm eff}=8200$~K and $\log{g}=4.0$, to be $B=2.65\pm0.15$ from fig.~2 in \cite{plac19}. The secondary gives practically no contribution to the effect so we neglected it. The theoretical Doppler beaming accounts for about half of the observed sinusoidal pattern of the residuals. We subtracted that effect from the TESS light curve and repeated the fitting procedure, this time obtaining a much better agreement with the out-of-eclipse light changes and also a slight improvement in the primary eclipse. Finally we adjusted the orbital eccentricity and the argument of periastron, which enabled a further decrease of the residuals in both the light and radial velocity curves.

Solutions to the radial velocity curves are fully satisfactory, with an $rms$ of the radial velocity residuals of 83~m~s$^{-1}$ for the primary but much a larger value of $\sim660$~m~s$^{-1}$, for the secondary. Although the primary is a mid A-type star, it is also metallic-lined. This allows a relatively precise determination of its radial velocitities despite the large rotational broadening of the lines ($v\sin{i}\sim40$~km~s$^{-1}$). The secondary, on the other hand, is practically invisible in the spectra, even in those with the highest S/N of 140. This causes the low precision of individual radial velocities and the large $rms$ for the secondary.

The primary is slightly distorted with a tidal deformation of 1.0\%. We estimated the spectral types as A4\,V + early K-type dwarf.

\subsubsection{CQ Ind}
This system consists of two solar-type stars but with components significantly different in physical appearance: the slightly evolved primary is a little more massive and hotter than the unevolved secondary. At first we used in our analysis only the TESS photometry from sector 27 combined with HARPS velocimetry. Although the secondary eclipse is almost perfectly placed at orbital phase 0.5 the system possesses significant eccentricity: the secondary eclipse is nearly twice as long as the primary eclipse. The primary eclipse has a short phase of totality when the secondary transit over the primary stellar disc, while the secondary eclipse is partial because it occurs when the stars are further apart from each other. It was quickly clear that apsidal motion is also significant and is influencing the combined fit because the radial velocity data cover more that 4~yr. In order to properly account for the apsidal motion we included in the analysis the first two eclipses observed by TESS in sector 1. The timespan between these two sectors is about 700~d ($\sim$80 orbital cycles), but with the precision of the TESS photometry it is possible to determine the rate of apsidal motion.

We did not solve the photometry and velocimetry simultaneously, but instead applied an iteration scheme: using photometry we determined the apsidal motion rate ($d\omega/dt$), the position of the orbit ($i$, $\omega$, $e$), the relative sizes of stars ($r_{1,2}$) and the relative temperatures ($T_{1,2}$), then we solved the radial velocity curves to get the semimajor axis of the system $a$, the mass ratio $q$ and revised values of $\omega$ and $e$. We repeated these steps until we obtained a satisfactory consistency in the orbital parameters ($\omega$ and $e$) derived from the photometry and velocimetry separately. Fig.~\ref{fig:light} shows the best light curve fit for CQ Ind where the points denote data only from sector 27. The rate of apsidal motion is slow, $2.9\times10^{-4}$ deg cycle$^{-1}$, corresponding to an apsidal period of $\sim$30\,000 years.

We adjusted the third light as its inclusion reduces the residuals during the eclipses, however its value is slightly negative ($l_3=-0.006$). We also adjusted the albedo on the secondary component and obtained $A_2 = 0.42\pm0.06$, in agreement with the expected value for fully convective stellar atmospheres. The estimated spectral type is F6\,V + F8\,V, in good agreement with the F7\,V reported by \cite{hou78}. The metallicity is solar and both components rotate synchronously. The radial velocity solution presented in Fig.~\ref{fig:rv} shows very small residuals with an $rms$ of 32~m~s$^{-1}$ for the primary and 25~m~s$^{-1}$ for the secondary.

\begin{sidewaystable*}
\tiny
\caption{\tiny Model parameters from the Wilson-Devinney code}\label{tab_par_orb}
\centering
\begin{tabular}{@{}rlccccccccccccl@{}}
\hline\hline
&&&&&&&&&&&&&&\\
ID & $P_{\rm orb}/T_0$ & $q=\frac{M_2}{M_1}$ & $a$ & $\gamma$ & $e$ & $\omega$ & $\frac{d\omega}{dt}$ & $K$ & $i$ & $T_{\rm eff}$ & $\Omega$ & $r$ & $\frac{L_2}{L_1}$ & $l_3$\\
& (days/BJD) & & ($R_\odot$) & (km s$^{-1}$) & & (deg) & (deg year$^{-1}$) & (km s$^{-1}$) & (deg) & (K) & & & & \\
&&&&&&&&&&&&&&\\
\hline
 GW~Eri p & 3.6586647(8)   &  0.9906(19) & 15.272(15) & 32.32(13) & 0 & - & - &  104.51(14) & 83.933(51) &  8270     & 9.353(71) &  0.11971(101) & 0.9417 & 0.028(3) \\
 s		    & 2459153.1534	& 		    &			& 32.32(13) &	 &     &   & 105.49(15) & 			& 8180(2) & 9.436(71) & 0.11760(98) & 		&		\\
HD~32129 p & 16.4120349(7) & 0.6945(7) & 37.407(26) & 4.867(15) & 0.4374(7) & 61.20(2) & 0.0042(4) &  52.546(31)	& 88.797(12) & 6725 & 21.212(52) & 0.05006(13) & 0.1724 & 0.022(2) \\
s		&2459481.5181	&		&		& 5.370(10)	& 			& 		&		& 75.656(68) 	&		& 5775(3) & 28.24(13) & 0.02640(13) & 0.1571\tablefootmark{a} & 0.009(4)\tablefootmark{a}  \\
UW~LMi p & 3.87431667(14) & 0.9807(9) & 13.7086(63) & $-$33.930(24) & 0 & - & - 	& 88.480(49) & 86.621(44) & 5970   & 11.21(11) &  0.09778(100) & 0.8804\tablefootmark{b} & 0\tablefootmark{b} \\
s		   & 2450854.7741  &		 & 			& $-$33.897(26) & 	&	&	& 90.223(65) & 			& 5935(2) & 11.56(14) & 0.09300(118) & 0.8645\tablefootmark{c}  & 0\tablefootmark{c} \\
QR Hya p & 5.0058710(4) & 0.9701(11) & 15.9370(87) &  10.534(18) & 0.0001(1) & 217(11) & - & 79.128(67) & 86.087(7) & 5880 & 13.383(39) & 0.08059(25) & 0.8378 & 0 \\
s		& 2458559.4578   & 			&        		& 10.603(17) &		&		&	 &  81.567(57) & 		&  5801(2) & 13.873(58) & 0.07547(34) &  & \\
V788 Cen p& 4.96637676(9) & 0.7881(14) & 18.6193(18) & $-$18.700(18) & 0 & - & - & 82.918(51) & 82.821(41) & 7820 & 7.613(22) & 0.14682(49) & 0.3126 & 0.021(2) \\
s		& 2458578.9929   & 			&        		& $-$18.919(24) &		&		&	 &  105.21(17)  & 		&  7406(3) & 10.046(58) & 0.08821(58) &  & \\
V338 Vir p & 5.9853360(16)& 1.1586(11) & 19.6047(89) & 2.760(13) & 0 & - & - & 88.572(50) & 84.747(33) & 6582(3) & 13.952(84)  &  0.07819(51) & 2.2116\tablefootmark{a} & 0\tablefootmark{a} \\
s		& 2457228.4125 & 		&		& 2.635(13)&	& 		&	& 76.446(55) &  	& 6425 &  10.535(38) & 0.12064(47) & & \\
V963 Cen p & 15.269303(7) &  0.9945(7) &  33.463(30) & $-$30.450(13)	& 0.4223(16)& 140.10(3) &  0.0065(5) & 60.920(35) & 87.255(32) & 5800 &   24.869(81) & 0.04320(15) & 0.9714 & $-$0.028(4) \\
s		& 2459356.3465 &				&		& $-$30.448(12)	&		&		&		& 61.257(24) &			&	5808(6) &25.15(11) & 0.04248(20) & 0.9731\tablefootmark{b} & 0\tablefootmark{b} \\
LX Mus p & 11.750601(2) & 1.0082(3) & 30.2789(46) & $-$4.860(11) & 0.1975(2) & 148.59(6) &  -   & 66.708(15) & 87.603(10) & 6525	& 23.77(16) & 0.04442(28) & 1.0941 & 0 \\
s		& 2459334.453	 &		 & 			& $-$4.881(11) &		&		&  		& 66.163(12) &			& 6556(2) & 23.11(12) &0.04610(23) & & \\
V362 Pav p & 2.7484368(5) & 0.4530(19)  & 11.655(32) & $-$0.83(6)   & 0.0014(4) & 283(16) & -& 66.555(81) & 84.304(35) & 8200 & 5.785(17) & 0.18817(61) &  0.0284 & $-$0.054(5) \\
s			& 2458672.6411 & 		&			& $-$0.06(25) &		&  &  & 146.93(58) &			&  4962(3)  & 7.542(37) &  0.07231(44) & & \\
CQ Ind p  & 8.9737116(2)	 & 0.8896(5)&  24.3265(78) & $-30.820$(12) &0.2764(5)& 89.66(1)& 0.0118(6)	 &  67.179(32)  & 89.159(10)&  6440 & 18.489(46) & 0.05795(16) & 0.5460 &$-$0.006(3) \\
s 		& 2459046.259	  & 			&        		& $-30.568$(11) &		&		&	 &  75.515(25)  & 		&  6122(3) & 20.646(93) & 0.04632(23) &  & \\
\hline
\end{tabular}
\tablefoot{\tiny Quoted uncertainties are the standard errors from the Differential Corrections subroutine combined with errors from
Monte Carlo simulations with the JKTEBOP code ver.~34.\\
In the ID column ``p'' refer to the primary and ``s'' to the secondary.  The meaning of the columns are: the observed orbital period (and epoch of the primary eclipse $T_0$ given below), the mass ratio, the total semimajor axis $a=a_1+a_2$, the apparent systemic velocity of each component, the orbital eccentricity, the longitude of periastron, the rate of apsidal motion, the radial velocity semiamplitude, the orbital inclination, the effective temperature, the Roche potential, the fractional radius, the light ratio in the TESS band and the amount of third light in the TESS band.\\
\tablefoottext{a}{\textit{Kepler} band}
\tablefoottext{b}{Str{\"o}mgren $y$ band}
\tablefoottext{c}{Str{\"o}mgren $u$ band}}
\end{sidewaystable*}

\begin{table*}
\centering
\caption{Physical parameters of the stars.}
\label{par_fi}
\begin{tabular}{@{}rccccccccc@{}}
\hline \hline
ID & $M$  & $R$ & $\log{g}$  & $T_{\rm eff}$  & $L$ & $\upsilon\sin{i}$ & $[{\rm M}/{\rm H}]$  & $\varpi_{\rm phot}$ & $E(B\!-\!V)$  \\
     & (M$_{\sun}$) & (R$_{\sun}$) & (dex) & (K) & (L$_{\sun}$) & (km s$^{-1}$) & (dex) & (mas) & (mag) \\
\hline
 GW Eri p & 1.7936(57) & 1.828(16) & 4.168(7) & 8370(82) & 14.1(6) & 25(1) & +0.51(15) & 11.62(22) &   0.001(5) \\
 s		 & 1.7768(54) & 1.796(15) & 4.179(7) & 8180(81) & 13.0(6) & 	 24(1) &	&				&		\\
HD 32129 p & 1.5388(36) & 1.8726(50) & 4.080(2) & 6710(60) &  6.40(23) & 1(1) &  +0.19(7) & 5.79(12) & 0.110(22) \\
s		 & 1.0687(20) & 0.9875(49) 	& 4.478(4) & 5760(97) & 0.97(7) & 2(2)	 & 		&		& \\
UW LMi p & 1.1627(18) & 1.340(14) & 4.249(9) & 6035(75) & 2.15(11)  &  17.4(7) & $-$0.10(6) & 9.55(18) & 0.005(5) \\
s		  & 1.1402(15) & 1.275(16) & 4.284(11) & 6000(72) &  1.90(10) & 17.1(8) & 			&		& \\
QR Hya p & 1.1002(18) & 1.2844(40) &4.262(3) &5925(75)  & 1.83(9) & 13.2(8) & $-$0.01(6)& 10.76(20) &0.003(3) \\
s		& 1.0673(19) & 1.2028(55) & 4.306(4) &5845(71) & 1.52(8) & 12.3(9) & & & \\
V788 Cen p& 1.9621(70) & 2.733(10) & 3.858(3) & 7820(105) &25.2(1.4) & 20.3(7) & +0.5(2) & 11.11(20) & 0.006(4) \\
s		  & 1.5463(34) & 1.642(11) & 4.197(6) & 7405(120) &7.31(48) & 17(2) & & & \\
V338 Vir p & 1.3074(20) & 1.5329(99) & 4.183(6) & 6580(89) & 3.97(22) &  10(1) & $-$0.10(6) & 3.91(8) & 0.024(10) \\
 s		 & 1.5148(21) & 2.3651(93) & 3.871(3) & 6425(62) & 8.59(34) & 13.3(7) & & & \\
V963 Cen p & 1.0812(29) &  1.4456(52) &  4.152(3) & 5810(58) & 2.15(9) & 8.4(8) & $-$0.06(5) & 8.87(22) & 0.018(10) \\
s		& 1.0753(30) & 1.4215(68) & 4.164(4) & 5820(67) &  2.09(10) & 8.2(7) & 			&		& \\
LX Mus p 	& 1.3433(6) & 1.3450(85) &  4.309(5) & 6535(70) & 2.97(12)& 4(1) & +0.09(5) & 6.91(16) & 0.056(12) \\
s 		& 1.3544(7) & 1.3959(70) & 4.280(4) & 6565(64) & 3.26(13) & 5.1(8) &  & & \\
V362 Pav p & 1.936(18) & 2.1931(93) & 4.043(3) & 8200(70) & 19.6(7) & 39(1) & +0.02(15) & 6.72(13) & 0.016(10) \\
s			& 0.8767(51) & 0.8428(56) & 4.530(5) & 4950(200) & 0.39(6) & 20:  & 		  &	& \\
CQ Ind p & 1.2694(12) & 1.4097(39) & 4.243(2) & 6460(68) & 3.12(13) & 8.1(8) & $-$0.04(8) & 8.99(16) & 0.006(5) \\
 s 		& 1.1293(12) & 1.1268(56) & 4.387(4) & 6140(71) & 1.63(8) & 6(1) & & & \\
\hline
\end{tabular}
\end{table*}

\section{Comparison with previous studies}
\subsection{GW Eri}
The first spectroscopic orbit of the system was provided by \cite{AbL77} based on 30 medium-resolution spectra taken at the 2.1-m coud{\'e} spectrograph at Kitt Peak between 1970 and 1976. The authors noted the extreme similarity of both components. Their radial velocity semiamplitudes are in very good agreement with ours, as is their systemic velocity. Their identification of the components is the same. The reference time $T_{\rm max}$ of the secondary quadrature provided by \cite{AbL77} is in perfect agreement with our result, with a difference of only $0.0003\pm0.011$ day, which indirectly shows that significant period changes are unlikely.

A combined light- and radial velocity solution was presented by \cite{Ver06}. They secured 22 high-resolution \'echelle spectra with the EBASIM spectrograph at the 2.1-m telescope at Complejo Astron{\'o}mico El Leoncito and CCD photometry in the $V$-band using the Helen Sawyer Hogg 0.6-m telescope. They used the WD code (version not specified) to derive the physical parameters of the components. Their masses are perfectly consistent with ours, while their radii are consistent with those from our unconstrained light curve solution. The reference time of the primary eclipse differs from our reference time by only $0.0004\pm0.0004$ day.

\subsection{UW LMi}
\cite{cla01}, in the end of a section devoted to UW~LMi, gave a reference to a forthcoming paper by Helt et al.\,containing a detailed analysis of this system. However, the paper was never published. A quantitive description of the system given by \cite{cla01} is in agreement with our results. \cite{Gri01} presented many more details of the system. He reported that the CORAVEL dips in his radial-velocity traces are slightly deeper for the primary star than for the secondary, and that the resulting difference in $V$-band magnitude between the components is about $0.15\pm0.05$ mag. Such a difference corresponds to a light ratio of $0.87\pm0.04$, which is consistent with our findings (Section~\ref{uwlmi}). We do not confirm his finding that the variance of the radial velocities of the primary is larger than that of the secondary's. In fact, as he already suggested, that was indeed due to a statistical fluke and our measurements show a larger radial velocity $rms$ of the secondary star, as one would expect. We found the systematic velocity about 1.5~km~s$^{-1}$ higher than Griffin's value, but we attribute this difference entirely to a zero-point instrumental shift between HARPS and CORAVEL. Our radial velocity semiamplitudes are in good agreement with Griffin's values though an order of magnitude more precise. We also derived the same rotational velocities of the components.

A combined analysis of \textit{Hipparcos} photometry and Asiago \'Echelle velocimetry was presented by \cite{mar04} and it shows a familiar picture: two components similar to each other. However, the precision of the determined parameters is much lower than from our work, and furthermore they are not consistent with our results. The primary star, which is the more massive and larger component, was assigned by them to be the secondary, less massive and smaller star. Their masses and radii differ from ours by $5\sigma$ and $2\sigma$, respectively. A much better agreement occurs for the orbital inclination and the orbital period. \cite{mar04} reported also unexpectedly high surface temperature for both stars ($T_{\rm eff}\approx 6500$ K) based on the strength of the Paschen 14 line relative to the Ca\,II triplet. We find much lower temperatures in accordance with the spectral type of UW LMi, the physical parameters of the stars, and the mass-luminosity relation for main-sequence stars \citep[$\sim 6000$ K;][]{eke15}.

\subsection{V788 Cen}
\cite{Cou74} presented a $V$-band light curve of the system showing two equal, shallow minima spaced by half an orbital period. However no analysis based on this light curve was published. The ephemeris given by \cite{Cou74} is in extremely good agreement with ours: the difference between the predicted and measured time of the primary eclipses in the TESS light curve is less than 1 minute, although the epochs differ by 47 years. Thus period changes in the system are unlikely.

\subsection{V963 Cen}
Preliminary results from the analysis of photometry and CORAVEL radial velocities reported by \cite{cla01} showed two nearly identical components with masses $\sim$1~M$_\odot$ on eccentric orbit. A detailed study of this system was announced but never published. A more comprehensive analysis of V963 Cen was presented by \cite{syb18}. They derived a very precise spectroscopic orbit in order to study the Rossiter-McLaughlin effect, and supplemented this by rather low-precision photometric parameters derived from an analysis of ASAS-3 data \citep{poj02}. The reported radial velocity semiaplitudes $K_{1,2}$ are practically identical to ours and the resulting masses $M_{1,2}$ are the same to within the errors. However, the errors quoted by \cite{syb18} for the semiaplitudes and masses, are surprisingly small. We suspect that by fixing the orbital eccentricity to $e=0.4217$ in their fit they artificially assumed a zero uncertainty on $e$. In our solution an uncertainty in the eccentricity is an important contribution to the error budget of $K_{1,2}$ and especially $M_{1,2}$. In fact our errors for the mass measurements are {\it dominated} by the error in $e$. If we assume an eccentricity with a standard and unrealistically small error from the WD code ($e=0.4223\pm0.0002$) that leads to smaller uncertainties in our $K_{1,2}$ and uncertainties in $M_{1,2}$ that are smaller by a factor of three -- in rough agreement with uncertainties reported by \cite{syb18}. However, their eccentricity is in perfect agreement with our value derived from velocimetry alone, although such a value of $e$ results in a relatively poor fit to the TESS light curve (see Section~\ref{v963cen}).

\section{The properties of new systems versus the surface brightness -- colour relation}\label{sec:sbcr}
We checked how the components of the ten systems in this work appear on a SBCR plot. We chose a standard relation between the surface brightness in the $V$-band and the $(V\!-\!K)$ colour. We expressed $K$ magnitudes in the 2MASS system \citep{skru06}. The light ratios of the components in Johnson $V$ and 2MASS $K$ bands which were extrapolated from our WD models and used to obtain individual intrinsic magnitudes are given in Table~\ref{tab:lratio}. Inspection of the positions of components against the SBCR gives immediate indications about any peculiarities, e.g. stars significantly above the mean SBCR are in most cases unrecognised multiple stellar systems or may have an incorrect value for third light. On the other hand a position significantly below may signify problems with adopted magnitudes e.g.\ a magnitude is calculated based on observations taken during eclipse without a correction for the light diminution. Another possibility is that the parallax is biased toward too large a value. Also, systems with a large reddening due to interstellar extinction could shifted away from SBCR if the reddening is not correctly accounted for.

The surface brightness parameter $S_{\!V}$ was calculated for our stars using equation 5 from \cite{hin89}:
\begin{equation}
\label{equ:sb}
S_{\!V} = 5 \log{\theta_{\rm LD}} + V_0,
\end{equation}
where $V_0$ is the intrinsic magnitude of a star in the $V$ band and $\theta_{\rm LD}$ is the limb-darkened angular diameter expressed in milliarcseconds. The angular diameters were calculated using:
\begin{equation}
\theta_{\rm LD} =  9.301 \times 10^{-3}\, R\, \varpi_{Gaia/EDR3},
\end{equation}
where $R$ is the stellar radius expressed in nominal solar radii $\mathcal{R}_\odot$ \citep{prsa16}.

\begin{figure}
\hspace*{-0.5cm}
\includegraphics[angle=0,scale=0.65]{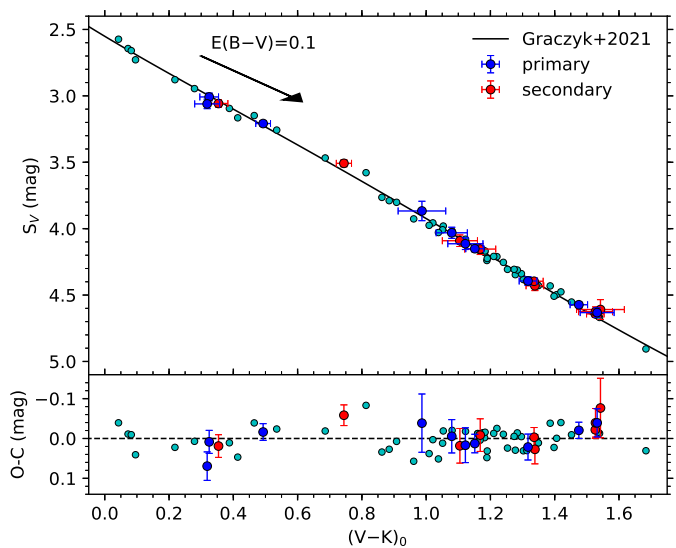}
\caption{The $V$-band surface brightness versus $(V\!-\!K)$ colour relation. The components of the ten eclipsing binaries in this work are shown by blue (primary) or red (secondary) points. 
The components of HD 32129 have the largest errorbars. Green points are data from \cite{gra21}.}\label{SBCR}
\end{figure}

We corrected the magnitudes of the HD~32129 system due to the presence of a putative K6/7\,V companion star. For all ten systems 2MASS magnitudes were taken outside eclipse, so there is no need to correct them for light loss. We adopted parallaxes from \textit{Gaia} Early Data Release 3 \citep{gaia20}. We did not use any corrections to parallaxes \citep[e.g.][]{lind20a,lind20b} because the systems are relatively close to us and the largest correction (V338 Vir) amounts to only 0.5\% of the parallax itself. From the sample one system, HD 32129, has the Gaia \verb"RUWE" parameter (the Renormalised Unit Weight Error)\footnote{\texttt{https://gea.esac.esa.int/archive/documentation/GDR2/Gaia\\\_archive/chap\_datamodel/sec\_dm\_main\_tables/ssec\_dm\_ruwe.html}} greater than 1.4 and also the largest fractional error of the parallax. Fig.~\ref{SBCR} shows the positions of the eclipsing binary components on the $V$-band surface brightness versus ($V\!-\!K$) colour diagram. The data are taken from our previous work \citep{gra21} and errorbars were suppressed in order to make the present sample clearly identifiable. The largest errorbars are those of the HD 32129 system. Practically all components lie very close to the SBCR derived from other eclipsing binary stars \citep{gra21} and the largest offsets are smaller than 2$\sigma$. New calibrations of the SBCRs utilizing the present, additional sample of stars are envisioned for a separate paper.

\section{Final remarks} \label{fin}
We present a detailed analysis of ten well-detached eclipsing binary stars. For the first time for all those systems, very precise and accurate astrophysical parameters were determined, including masses, radii, temperatures and metallicity. The high precision of the determined parameters makes these systems valuable for testing stellar evolution models. One system, GW Eri, is a visual triple system. Another, HD 32129, is a suspected triple system with a tertiary close to the main binary system. In principle all systems, with the possible exception of HD 32129, are useful for recalibration of the SBCRs based on {\it Gaia} EDR3 and later releases.

At least 30 more suitable DEBs lying within 250~pc of the Sun are expected to be analysed by our team in the near future. These systems, in combination with those with published detailed analysis, will be used to discuss issues such as the gravity and metallicity dependence of SBCRs. They will also be used for new calibrations of the stellar surface temperature versus colour relations.

\begin{acknowledgements}
We thank an anonymous referee for improvements in the text of this paper.
This work has made use of data from the European Space Agency (ESA) mission
{\it Gaia} (\url{https://www.cosmos.esa.int/gaia}), processed by the {\it Gaia}
Data Processing and Analysis Consortium (DPAC,
\url{https://www.cosmos.esa.int/web/gaia/dpac/consortium}). Funding for the DPAC
has been provided by national institutions, in particular the institutions
participating in the {\it Gaia} Multilateral Agreement.
\\
We are grateful to J.V.~Clausen, B.E.~Helt, and E.H.~Olsen for making their
unpublished $uvby$ photometric data available to us.
\\
The research leading to these results has received funding from the European Research Council (ERC)
Synergy "UniverScale"  grant financed  by  the  European  Union's  Horizon  2020  research and  innovation  programme  under  the  grant  agreement  number  951549,
 from the National Science Center, Poland grants MAESTRO UMO-2017/26/A/ST9/00446 and
BEETHOVEN UMO-2018/31/G/ST9/03050. We acknowledge support from the IdP II 2015 0002 64 and
DIR/WK/2018/09 grants of the Polish Ministry of Science and Higher Education.
\\
The research was based on data collected under the ESO/CAMK PAN - OCA agreement at the ESO Paranal Observatory.
\\
W.G. also gratefully acknowledges
support from the ANID BASAL project ACE210002.

W.G.\ also gratefully acknowledges financial support for this work from the BASAL Centro de Astrofisica y Tecnologias Afines
BASAL-CATA (AFB-170002), and from the Millenium Institute of Astrophysics (MAS) of the Iniciativa Cientifica Milenio del Ministerio de Economia,
Fomento y Turismo de Chile, project IC120009. 
\\
A.G. acknowledges support from the ANID-ALMA fund No. ASTRO20-0059 and
MT acknowledges financial support from the Polish National Science Center
grant PRELUDIUM 2016/21/N/ST9/03310.
 \\
This research has made use of the VizieR catalogue access tool, CDS,
 Strasbourg, France (DOI : 10.26093/cds/vizier). The original description
 of the VizieR service was published in 2000, A\&AS 143, 23.
 \\
 Based on observations made with ESO Telescopes at the La Silla Paranal Observatory under programmes 082.D-0499, 083.D-0549, 084.D-0591, 085.C-0614, 085.D-0395, 086.D-0078, 091.D-0469, 092.D-0363, 094.D-0056, 095.D-0026, 097.D-0150, 099.D-0380, 0100.D-0273, 0100.D-0339, 0101.D-0697, 0102.D-0281, 105.2045.002, 105.20L8.002, 106.20Z1.001, 106.20Z1.002, 108.21XB.001, 190.D-0237 to PIs: G.P., W.G. and D.G.; also 087.C-0012(A) to PI Krzysztof He{\l}miniak, 089.C-0415(A) and 094.C-0428(A) to PI Rafael Brahm. 

 \\
We used the {\it uncertainties} python package.

\end{acknowledgements}


\label{lastpage}

\end{document}